\documentclass[twocolumn,showpacs,superscriptaddress,preprintnumbers,amsmath,amssymb,epsfig,floatfix,prl]{revtex4-1}
\usepackage{epsfig}
\usepackage{graphicx}
\usepackage{dcolumn}
\usepackage{color}
\usepackage{natbib}  
\definecolor{blue}{RGB}{0,71,0}
\usepackage{bm}
\usepackage{tabularx}
\newcolumntype{L}[1]{>{\raggedright\arraybackslash}p{#1}}
\newcolumntype{C}[1]{>{\centering\arraybackslash}p{#1}}
\newcolumntype{R}[1]{>{\raggedleft\arraybackslash}p{#1}}
\begin{document}
\title{Transition Metal and Vacancy Defect Complexes in Phosphorene: A Spintronic Perspective}
\author{Rohit Babar}
\affiliation{Department of Physics, Indian Institute of Science Education and Research, Pune 411008, India}	
\author{Mukul Kabir}
\email{Corresponding author: mukul.kabir@iiserpune.ac.in} 
\affiliation{Department of Physics, Indian Institute of Science Education and Research, Pune 411008, India}
\affiliation{Centre for Energy Science, Indian Institute of Science Education and Research, Pune 411008, India}
\date{\today}

\begin{abstract} 
Inducing magnetic moment in otherwise nonmagnetic two-dimensional semiconducting materials is the key first step to design spintronic materials. Here, we study the absorption of transition-metals on pristine and defected single-layer phosphorene, within density functional theory. We predict that increased transition-metal diffusivity on pristine phosphorene would hinder any possibility of controlled magnetism, and thus any application. In contrast, the point-defects will anchor metals, and exponentially reduce the diffusivity.  Similar to other two-dimensional materials, metals bind strongly on both pristine and defected phosphorene, and we provide a microscopic description of bonding, which explain the qualitative trend with increasing number of valence electrons.  We further argue that the divacancy complex is imperative in any practical purpose due to their increased thermodynamic stability over monovacancy. For most cases, the defect-transition metal complexes retain the intrinsic semiconduction properties, and also induce a local magnetic moment with large exchange-splitting and spin-flip energies, which are necessary for spintronic applications. Specifically, the V/Mn/Fe absorbed at the divacancy have tremendous promise in this regard. Further, we provide a simple microscopic model to describe the local moment formation in these transition metal and defect complexes. 

\end{abstract}
\maketitle

Atomically thin two-dimensional (2D) materials have been a subject of intense research due to the unique physical properties originating from 2D confinement of electrons, and concurrent applications in electronic and optoelectronic devices.~\citep{10.1038/nmat1849, 10.1038/nnano.2010.89, 10.1038/nature11458, 10.1021/nn400280c}  In this regard, graphene has attracted much attention due to very high carrier mobility, $\sim$ 10$^5$ cm$^{2}$V$^{-1}$s$^{-1}$.\citep{10.1038/nnano.2010.89,PhysRevLett.100.016602} However, graphene is semimetal,\citep{10.1038/nmat1849, 10.1038/nnano.2010.89, 10.1038/nature11458,PhysRev.71.622} and the lack of a band gap severely limits its application in field-effect transistor due to insufficient on-off current ratio.  Thus, although without much success, extraordinary efforts have been given to induce a semiconducting gap without compromising the exceptional carrier mobility by structural modification of graphene\citep{PhysRevLett.97.216803,10.1038/nmat2003, 10.1038/nmat2710} or by external perturbations such as electric field and strain.\citep{PhysRevLett.102.256405,PhysRevB.81.081407} 
In contrast, other 2D materials such as chalcogenides,~\citep{10.1021/nn400280c,10.1038/nnano.2012.193} oxides,~\citep{ADMA201103241} and III-VI semiconductors~\citep{ADMA201201361} posses sizeable band gaps around 1$-$2 eV.  Many transition metal dichalcogenides monolayers are natural direct band gap semiconductors.~\citep{10.1038/nnano.2012.193}  For example, monolayer MoS$_2$ has a direct band gap of $\sim$ 1.8 eV, which corresponds to a device on-off current ratio of 10$^8$.~\citep{10.1038/nnano.2010.279} Although the carrier mobilities are much lower (200 cm$^2$V$^{-1}$s$^{-1}$) than in graphene, it is still remarkably high compared to the thin-film semiconductors.~\citep{10.1038/nnano.2010.279} 

Recently, single layer of phosphorus arranged in a puckered hexagonal lattice, phosphorene, was isolated by mechanical exfoliation from bulk black phosphorus,~\citep{10.1038/nnano.2014.35,10.1021/nn501226z} which is a layered van der Waals material, much like bulk graphite.  Although the stability has been a challenge as thin phosphorene films degrade rapidly under ambient conditions, air-stable monolayers have been fabricated recently.~\citep{NatureCommun10450}
Inside a single phosphorene layer, each atom is covalently bonded with three adjacent atoms, and thus three of the valence electrons of phosphorus are consumed in bonding. Consequently, phosphorene is a 2D semiconductor with a direct band gap of  1.45$-$2.05 eV,~\citep{10.1021/nn501226z,10.1021/nl502892t} which is significantly larger than the bulk counterpart, $\sim$ 0.2.~\citep{10.1038/nnano.2014.35} This semiconducting gap of single-layer phosphorene correspond to a large on-off current ratio of 10$^5$, which is essential for any device applications.~\citep{10.1038/nnano.2014.35}
The field-effect carrier (hole) mobility is found to be thickness dependent, and reported to be as high as 1000  cm$^2$V$^{-1}$s$^{-1}$ at room temperature.~\citep{10.1038/nnano.2014.35,10.1021/nn501226z}  Further, the intrinsic band gap and mobility can be easily tuned by layer thickness~\citep{10.1038/nnano.2014.35,10.1021/nn501226z,10.1038/ncomms5475} and external stress.~\cite{PhysRevLett.112.176801} The intrinsic anisotropy in electronic, optical, thermal and mechanical properties make phosphorene an exciting material with promising applications in (opto) electronics, thermoelectronics, and spintronics.~\citep{10.1038/nnano.2014.35,10.1021/nn501226z,10.1021/nl502865s}    


Inducing magnetism in otherwise non-magnetic 2D materials could potentially facilitate the application of these materials in spintronic devices and magnetic recording media.~\citep{PhysRevB.77.195434, PhysRevLett.102.126807, 10.1038/nnano.2014.214} The quest of inducing local magnetic moment is the obvious first step in this regard, followed by inducing magnetic ordering, and ability to control and tune it  by gating, doping and fuctionalization. Transition metal (TM) doping is the conventional way to induce magnetism in such nonmagnetic 2D materials, and has been already deployed for graphene~\citep{PhysRevB.77.195434, PhysRevLett.102.126807,10.1038/nnano.2014.214} and phosphorene.~\citep{10.1021/acs.jpcc.5b01300, 10.1021/jp511574n, 10.1021/jp5129468} Interestingly,  cobalt adsorbed pristine phosphorene is found to show magnetic ordering, which could be further manipulated by gating.~\citep{PhysRevB.91.155138} Moreover, room-temperature ferromagnetism is predicted in this system. The magnetic ordering in such metal doped 2D systems and the concurrent Curie temperature are dictated by the relative distance and orientation between the doped TMs.~\citep{PhysRevLett.108.187208}  However, the magnetism may not be controllable in such systems due to very high TM diffusivity (resulting from their low activation barrier) on pristine 2D surface.~\citep{PhysRevLett.102.126807,PhysRevB.77.195434} Such high TM diffusivity will severely affect their spatial distribution on pristine 2D surface, resulting in uncontrolled magnetism. Therefore, it is desirable to decorate TMs at the intrinsic vacancy defects, which will anchor the TMs, and concurrently restrict their diffusion.~\citep{PhysRevLett.102.126807, SMLL200700929}  With the recent experimental advent, such vacancy defects and defect-TM complexes can be created with atomic precision that has been demonstrated already for graphene.~\citep{10.1038/nmat1996, 10.1021/nl2031629} 


Here we report the electronic and magnetic properties of 3$d$ transition-metals (TMs) absorbed on the pristine and defected single-layer phosphorene, and discuss our results in the context of spintronic applications. The present first-principles calculation is extensive; all the observed trends in electronic and magnetic properties are explained within microscopic models, and the important parameters for spintronic applications such as exchange-splitting and spin-flip energies are calculated. The TM diffusion on the pristine and defected surface has been studied, and we argue that TM doped pristine phosphorene is not suitable for any practical purpose due to very high TM diffusivity. In contrast to grapheme, we find that these vacancy defects are electronically inactive, and TM-vacancy complexes exhibit a diverse electronic and magnetic properties.

{\color{blue}\bf Computational details}. 
We use the spin-polarized density functional theory (DFT)~\citep{PhysRevB.47.558,PhysRevB.54.11169} within the projector augmented wave formalism~\citep{PhysRevB.50.17953} with a plane-wave cut-off of 500 eV. The exchange correlation is described with Perdew-Burke-Ernzerhof (PBE) form of generalized gradient approximation,~\citep{PhysRevLett.77.3865} if not explicitly specified.  Few of the calculations are repeated with Heyd-Scuseria-Ernzerhof (HSE06) hybrid functional.~\citep{jcp1.1564060} All calculations are carried out using a  7$\times$5 phosphorene supercell along the zigzag and armchair directions, respectively, and a vacuum layer of 15 \AA~ is used. Reciprocal space integration is carried out using 4$\times$4$\times$1 Monkhorst-Pack k-point sampling. All structures are optimized until all the forces on each atom are less than 0.01 eV/\AA.  Transition-metals are absorbed on the pristine, and defected single-layer phosphorene, and the present calculation refer to a very dilute limit of adatom/doping, 0.7\%. The DFT is coupled with climbing-image nudged elastic band (CINEB) method~\citep{1.1329672} to calculate the activation barrier for TM diffusion. The integrity of the minimum energy path and thus of the first-order transition state is confirmed by one-and-only-one imaginary phonon frequency.

{\color{blue}\bf Black phosphorus and single-layer phosphorene}. 
We start our discussion with bulk black phosphorus (BP) and single-layer phosphorene. The BP has orthorhombic lattice with space group $Cmca$, where covalently bonded puckered honeycomb layers are stacked together by weak van der Waals (vdW) interaction.  Calculated lattice parameters ($a$=3.34, $b$=4.46, and $c$=10.72 \AA), within dispersion corrected non-local optB88-vdW functional,~\citep{PhysRevLett.92.246401, PhysRevB.83.195131}  are in good agreement with experimental values,~\citep{10.1107/S0365110X65004140} and previous calculations.~\citep{10.1038/ncomms5475,10.1021/nn501226z} The BP has a direct band gap at the Z point of the first Brillouin zone, which is calculated to be 0.15 eV within the PBE functional. The modified Becke-Johnson exchange potential in combination with correlation within local (spin) density approximation (mBJ)~\citep{10.1063/1.2213970,PhysRevLett.102.226401} yield a larger band gap of 0.31 eV. These compare well with the predicted experimental gap, 0.2$-$0.35 eV.~\citep{10.1038/nnano.2014.35,PhysRev.92.580,10.1063/1.1729699,Maruyama198199} and previous theoretical calculations.~\citep{10.1038/ncomms5475,10.1021/nn501226z,PhysRevB.89.201408}  Similar to graphite, the individual layers are held together via weak vdW interaction in black phosphorus, and thus it is easy to exfoliate single and few-layer phosphorene.~\citep{10.1038/nnano.2014.35,10.1021/nn501226z}. Calculated phosphorene lattice parameters, $a$=3.30, $b$=4.62 \AA, are comparable to the bulk counterpart, and also agree well with previous reports.~\citep{10.1021/nn501226z,10.1038/ncomms5475,PhysRevB.89.201408,10.1021/acs.jpcc.5b01300,10.1021/jp5110938} In the puckered honeycomb lattice, two half-layers are separated by 2.10 \AA~ distance, and have two inequivalent P--P bonds, 2.22 (in-plane) and 2.26 \AA~ (out-of-plane).  Single layer phosphorene is found to have semiconductive nature with a direct band gap, which is calculated to be 0.90 and 1.48 eV within different treatment of exchange-correlation functionals, PBE and mBJ, respectively (Supporting Information, Figure S1).  This is in agreement with previous theoretical calculations.~\citep{10.1038/ncomms5475,10.1021/acs.jpcc.5b01300,10.1021/jp5110938,10.1021/nl5021393}  In comparison, the spectroscopic measurements predict the gap to be 1.45$-$2.05 eV.~\citep{10.1021/nn501226z,10.1021/nl502892t}  It is well known fact that conventional DFT calculations tend to underestimate the gap, which could be improved via computationally expensive hybrid exchange-correlation functional or many-body perturbation (GW) approach.~\citep{10.1021/nn501226z,10.1038/ncomms5475,PhysRevB.89.201408} Indeed, the HSE06 hybrid functional improves the band gap to 1.63 eV  (Figure S1) In the case of single and few-layer phosphorene, although the conventional PBE functional underestimates the gap, it is important to note that the electronic band dispersion is similar with hybrid functional  or GW approach,~\citep{10.1021/nn501226z,10.1038/ncomms5475,PhysRevB.89.201408} where the conduction band is pushed up in energy with a very little effect on the valence band (Figure S1).  
First-principles calculations demonstrated that the band gap decreases with increasing layer thickness,~\citep{10.1038/ncomms5475} and can also be tuned by external strain.~\citep{PhysRevLett.112.176801} It is important to note in Figure S1 that the conduction band is more dispersive along the armchair (linear along $\Gamma \rightarrow$ Y) direction than in the zigzag (parabolic along $\Gamma \rightarrow$ X) direction, shown in Figure S1. This is responsible for the anisotropic electrical conductance, which has been experimentally observed.~\citep{10.1021/nn501226z, 10.1038/nnano.2014.35, 10.1038/ncomms5475}

\begin{figure*}[t]
\centering
\includegraphics[scale=0.14]{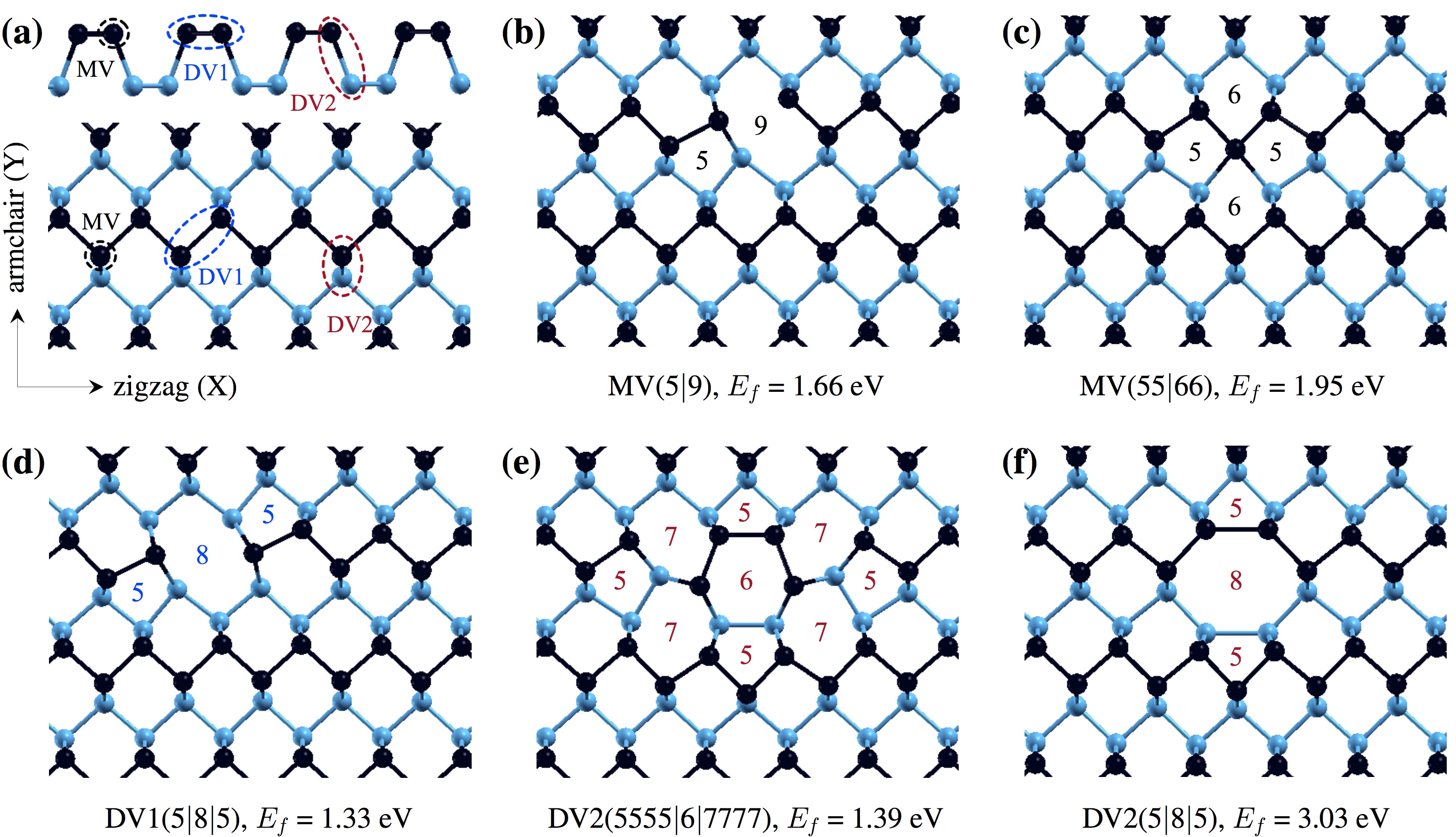}
\caption{Point-defect structures in single-layer phosphorene and their formation energy. (a) Side and top view of pristine phosphorene. Two half-layers are indicated in black and light blue color. The atoms are highlighted, which are removed to create point defects. Note that the DV can be created by removing two neighbouring atoms from the same or different half-layers, DV1 and DV2, respectively. (b)-(c) The MV defect structures, with MV(5$|$9) as the ground state. (d)-(e) Different DV structures. The DV1(5$|$8$|$5) is the ground state, which is thermodynamically more stable than MV.  Due to their very high formation energy, the other DV structures, DV2(5555$|$6$|$7777) and DV2(5$|$8$|$5) are much less probable. }
\label{figure1}
\end{figure*}

{\color{blue}\bf Monovacancy and divacancy.} 
Next we consider the bare point-defects, mono- and di-vacancies.  As-prepared 2D materials are prone to vacancy defects,~\citep{10.1021/nn102598m,10.1021/nl5021393,PhysRevLett.109.035503} which can also be created with atomic precision,~\citep{10.1038/nmat1996,10.1021/nl2031629} and may alter the intrinsic electronic, magnetic and chemical properties.~\citep{10.1021/nn102598m,PhysRevB.77.195428,PhysRevB.83.144115} In the spintronic context, it would be interesting to address if the intrinsic semiconducting property is retained as the point-defects are introduced in single-layer phosphorene.  Removing one P atom to create monovacancy (MV) in phosphorene [Fig.~\ref{figure1}(a)] leaves one dangling bond each on three adjacent P-sites before lattice relaxation. Following this, there are two distinct relaxation paths, which lead to two distinct MV structures, MV(5$|$9) and MV(55$|$66),~\citep{10.1021/nl5021393, 10.1021/acs.jpcc.5b06077,0957-4484-26-6-065705} shown in Fig.~\ref{figure1}(b)-(c). We find the Jahn-Teller distorted MV(5$|$9) [Fig.~\ref{figure1}(b)] to be the most probable monovacancy structure with 1.66 eV formation energy, which is 0.29 eV lower than the hypervalent MV(55$|$66) structure [Fig.~\ref{figure1}(c)]. The formation energy is calculated as $E_f = E_d - E_p + N\mu$, where $E_d$ and $E_p$ represent the energies of the fully relaxed system with and without the vacancy defect; $N$ is the number of removed atoms, and $\mu$ is the chemical potential of P calculated from pristine phosphorene. Similar MV structures have been observed for flat graphene~\citep{PhysRevB.68.144107, 10.1038/nature02817, PhysRevLett.95.205501,10.1021/nn401113r}  and slightly buckled silicene,~\citep{C3NR02826G} but the formation energies in these cases are much higher, $\sim$ 7.5 and 3 eV, respectively.~\citep{PhysRevB.68.144107,C3NR02826G}
Survival of one dangling bond in MV(5$|$9) induces a local magnetic moment of 1 $\mu_B$, and also generates a defect state 0.21 eV above the valence band maximum (VBM). Thus, bare MV(5$|$9) substantially reduces the intrinsic band gap of phosphorene (Figure S2). However, once the dangling bond is passivated (with atomic H in the present case), both the local moment and defect state disappear, and the intrinsic semiconducting gap (0.95 eV) is recovered (Figure S2), which increases to 1.71 eV within hybrid HSE06 functional (Figure S3).  

Divacancy in phosphorene can be created in two different ways by removing two neighboring P atoms from the same or different half-layers, which we classify as DV1 and DV2 type [Fig.~\ref{figure1}(a)]. Similar to the situation in graphene, the concurrent relaxation and reconstruction via Stone-Wales (SW) bond rotation may lead to a plethora of divacancy structures.~\citep{10.1021/acs.jpcc.5b06077,0957-4484-26-6-065705} We find the DV1(5$|$8$|$5), where the neighboring P atoms are removed from the same half-plane, to be the most stable defect structure with 1.33 eV formation energy [Fig.~\ref{figure1}(d)].  The other possibility of removing two neighboring atoms from two different half-layers relaxes to a different type of DV2(5$|$8$|$5) structure [Fig.~\ref{figure1}(f)] with a much higher formation energy (3.03 eV). The other DV classes, DV(555$|$777), and DV(5555$|$6$|$7777) can be reconstructed from either of the DV(5$|$8$|$5) structures by SW bond rotations, which is an activated process.  The DV2(5555$|$6$|$7777) shown in  Fig.~\ref{figure1}(e) is found to be close in energy with DV1(5$|$8$|$5) ground state, with 1.39 eV formation energy. This structure is derived from DV2(5$|$8$|$5) by two successive SW rotations to minimise the strain energy. In contrast, the DV(555$|$777) structures are found to be much higher in energy,~\citep{10.1021/acs.jpcc.5b06077,0957-4484-26-6-065705} and we did not consider them in the present study.  
Due to the complete quenching of dangling bonds, DVs are found to be nonmagnetic, and no mid-gap states are observed (Figure S4 and S5). Both DV1(5$|$8$|$5) and DV2(5555$|$6$|$7777) retain the intrinsic semiconducting nature of phosphorene with a band gap of 0.97 (Figure S4) and 1 eV, respectively. The calculated PBE and HSE06 gaps (Figure S5) for DV1(5$|$8$|$5) are close to the same calculated for the pristine single-layer phosphorene within the same level of theory.  Thus, the intrinsic semiconducting gap is robust against defects. 


Similar DV structures have been found in graphene~\citep{10.1021/nn102598m,PhysRevLett.95.205501} and silicene.~\citep{C3NR02826G} However, the qualitative stability of the DVs in these 2D materials is very different than the one predicted here for phosphorene. Moreover, similar to the case of MV, the DV formation energy in phosphorene is much smaller compared to graphene ($\sim$7.2 eV~\citep{0957-4484-26-6-065705}) and silicene (2.84 eV~\citep{C3NR02826G}). This indicates that point-defect formation is comparatively easier in phosphorene, and the reason is two fold.  The P--P bond is less stronger than the C--C bond in graphene and Si--Si bond in silicene. Secondly, due to the inherent puckered structure of the phosphorene lattice, the local strain relaxation is much easier.  

The remarkable reduction in DV formation energy, compared to that of the two isolated MVs, indicates that DVs are thermodynamically more stable than isolated MVs. The isolated MVs will coalesce into DV via diffusion, which requires a rather small activation barrier, 0.18 eV.~\citep{0957-4484-26-6-065705} Thus, DV is more important in the context of TM-vacancy complex, and for any practical purpose. However, these vacancy defects may appreciably alter the electronic structure of 2D materials by generating mid-gap states, which can act as sink for charge carrier and act as electron hole recombination centre. This may severely limit device performance in (opto)electronic applications. Unlike graphene and silicene,~\citep{10.1021/nn102598m,PhysRevB.77.195428,PhysRevB.83.144115} vacancy defects are found to be electronically inactive in phosphorene. We do not find any mid-gap states for hydrogenated MV and DV, and in both case the intrinsic semiconducting gap is retained (Supporting Information). The semiconducting properties are also found to be very robust against extended defects such as dislocations, and grain boundaries.~\citep{10.1021/nl5021393}

\begin{figure}[t]
\centering
\includegraphics[scale=0.15]{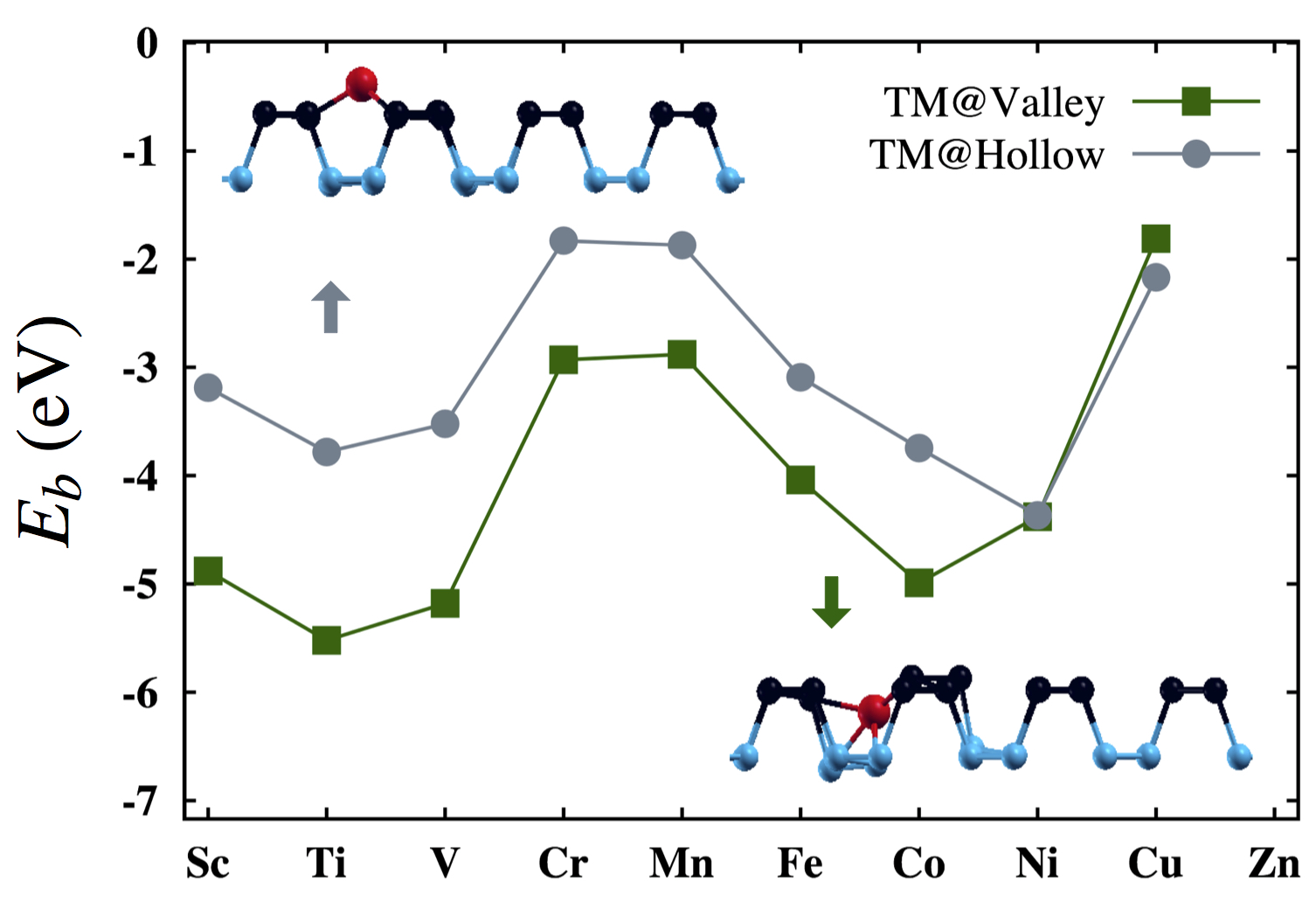}
\caption{Absorption energy of 3$d$ TM adatom on pristine phosphorene. Schematic illustration of adatom absorption at the hollow and valley sites are shown. For Sc--Co, a new absorption site TM@Valley is found to be thermodynamically more favourable than absorption at the hollow site, which has been proposed earlier.~\citep{10.1021/jp5110938,10.1021/acs.jpcc.5b01300,10.1021/jp5129468} The overall trends could be corroborated through COHP calculation. }
\label{fig:adatom-BE}
\end{figure}

{\color{blue}\bf Inducing local moment: TM on pristine phosphorene.} 
The conventional way to induce local magnetic moment in otherwise non-magnetic 2D materials is the absorption of TMs with localized $d$ electrons. This has been attempted for graphene~\citep{PhysRevB.77.195434} and recently for phosphorene.~\citep{10.1021/acs.jpcc.5b01300,10.1021/jp5129468,10.1021/jp5110938} Following these recent reports,~\citep{10.1021/acs.jpcc.5b01300,10.1021/jp5129468,10.1021/jp5110938} it was earlier believed that TMs prefer the hollow site, where it is covalently bonded with three P atoms, all from the `top' half-layer (Figure~\ref{fig:adatom-BE} and Figure S6 in Supporting Information).  Due to puckered structure, the TM atoms are relaxed away from the exact hollow site, resulting in two equivalent TM--P bonds with equal bond lengths, which is slightly smaller than the third bond. The height of the TM atoms above the phosphorene surface monotonically decreases along the 3$d$ series, which is commensurate with the corresponding TM covalent radius (Supporting Information).

In contrast, we predict a completely different absorption site with different bonding, electronic and magnetic properties. We find that TM prefers the valley site, where it is absorbed in between the two half-layers (inset of Figure~\ref{fig:adatom-BE}) and forms five/seven TM--P bonds (Figure S6 in Supporting Information). The TM binding energy at the valley site is substantially stronger (by $-$0.9 to $-$1.8 eV) compared to hollow site absorption (Figure~\ref{fig:adatom-BE}).  The binding (absorption) energy is calculates as $E_b = E_{\rm TM@P}-E_{\rm P}-\mu_{\rm TM}$, where $E_{\rm TM@P}$ and $E_{\rm P}$ is the energy of the system with and without the TM, $\mu_{\rm TM}$ is the energy of a isolated TM. For Ni and Cu, both the absorption sites are found to be equally preferable. Due to closed-shell electronic configuration, Zn binds either very weakly ($-$0.15 eV at the hollow site) or does not bind at all (valley site). Incorporation of van der Waals interaction in the calculation does not improve the situation for Zn. Overall the metal atoms bind very strongly with phosphorene, and the binding energy ranges between $-$1.81 to $-$5.52 eV at the valley site (Figure~\ref{fig:adatom-BE}). This is much stronger compared to the same on graphene, which is estimated in the range of $-$0.17 to $-$1.95 eV.~\citep{PhysRevB.77.195434} The stronger TM binding is due to the covalent TM--P bonding (Figure S6) originated from the lone-pair electrons of P atoms. To corroborate the overall trends, greater stability of valley site absorption and variation in binding energy along the 3$d$ series, we have calculated the TM--P projected crystal orbital Hamiltonian population (pCOHP)~\citep{doi:10.1021/jp202489s, doi:10.1021/j100135a014} and are shown in Figure S6 (Supplementary Information). The integrated pCOHP, which is a measure of bonding, is found to be much higher for TM@Valley than the TM@Hollow configuration. For example, IpCOHP for Ti@Valley and Cr@Valley is much higher ($-$7.83 and $-$6.44 eV, respectively) than the corresponding absorption at the hollow site ($-$4.88 and $-$2.85 eV, respectively).

\begin{figure}[t]
\centering
\includegraphics[scale=0.15]{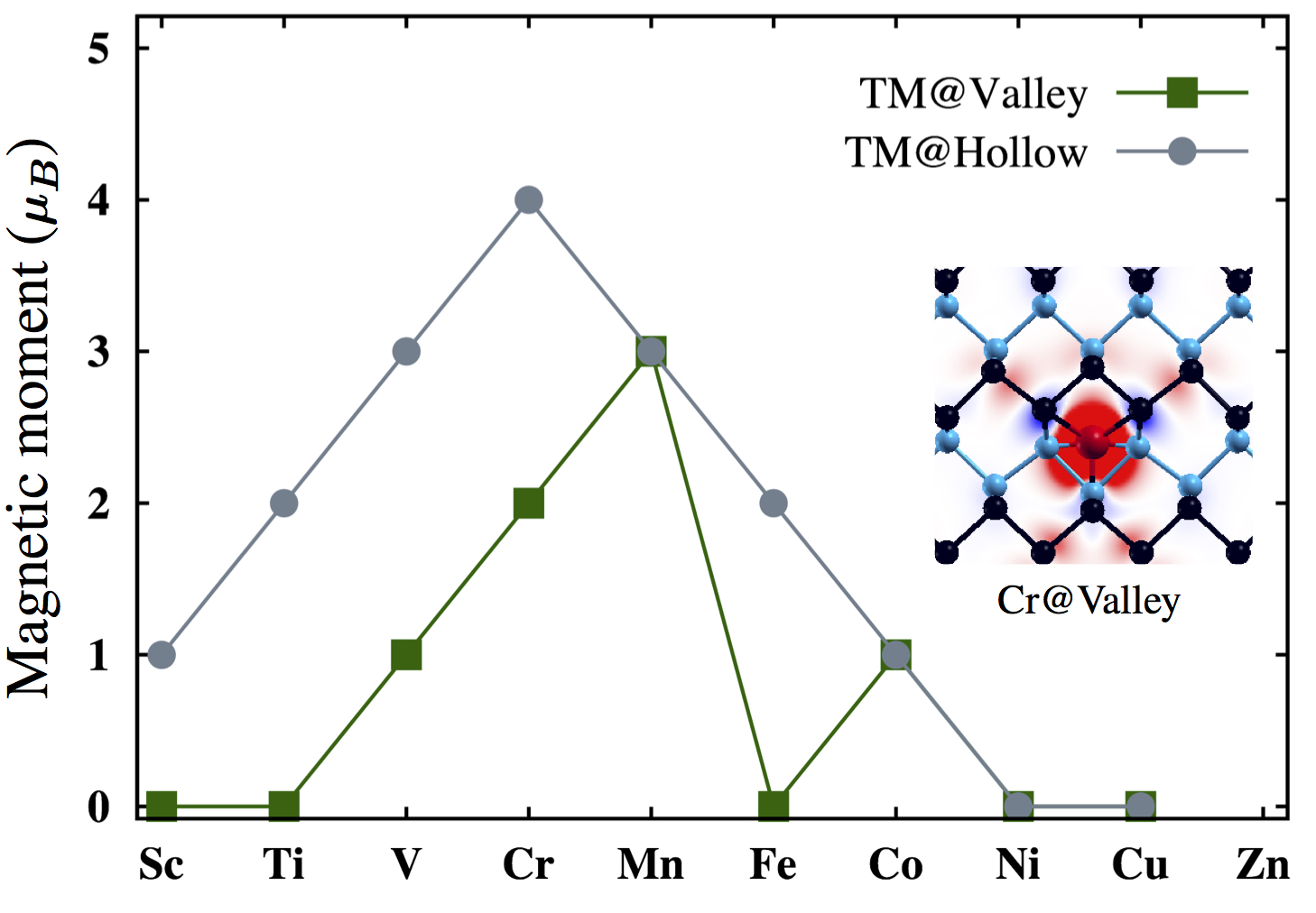}
\caption{Local magnetic moment at the TM site for TM absorption on the pristine phosphorene. In general, the TM moment is smaller in case of TM@Valley absorption, which is found to be the favorable absorption site. The trend along the 3$d$ series can be explained by simple electron counting. Representative magnetization density for the valley site absorption is shown in the inset.}
\label{fig:adatom-MM}
\end{figure}

\begin{figure*}[!t]
\centering
\includegraphics[scale=0.14]{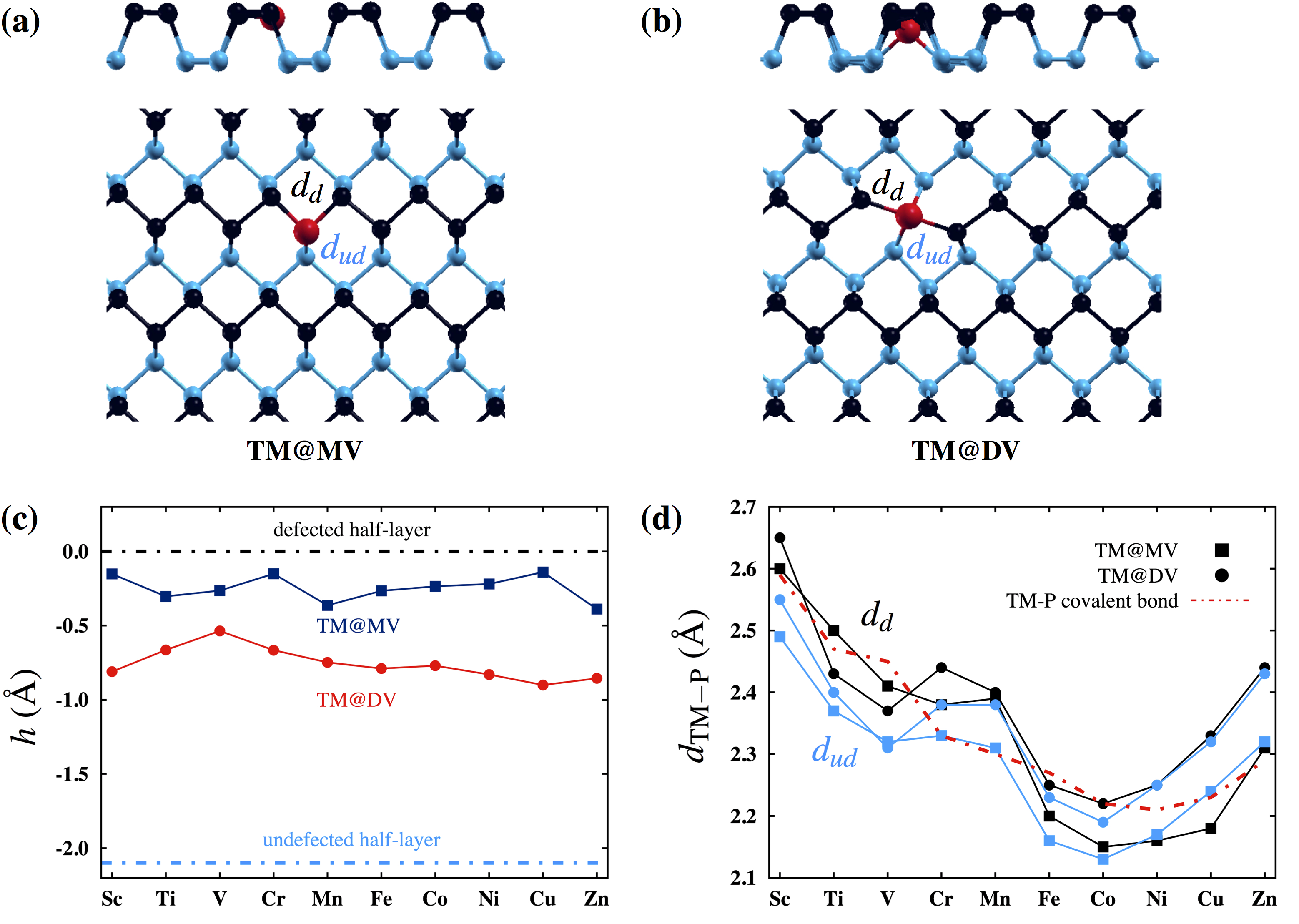}
\caption{Transition-metal (red ball) absorption at the point-defects in their ground state. Side and top view of (a) TM@MV, and (b) TM@DV. The TMs are absorbed in the valley, between the defected (black) and undefected (blue) half-layers. (c) Non-monotonous variation of the depth $h$ is observed along the 3$d$ series. The TM in TM@DV complex, move further away from the defected layer compared to the TM@MV complex. (d) The TM-P bonds ($d_{\rm TM-P}$), at the defected ($d_d$)  and undefected ($d_{ud}$) layers, are found to be covalent.  The red dotted line refers to the corresponding TM--P covalent bond lengths.~\citep{CHEM:CHEM200800987}}
\label{fig:structure-defect}
\end{figure*}

Now we turn our attention to the local moment induced by TM absorption. Calculated magnetic moments are shown in Figure~\ref{fig:adatom-MM} for both TM@Valley and TM@Hollow absorption. Although the absorption at the hollow site is not preferred,  calculated  moments agree well with previous predictions.~\citep{10.1021/acs.jpcc.5b01300,10.1021/jp5129468,10.1021/jp5110938} The magnetic moment at the valley site is smaller than the hollow site absorption for Ti-Cr, while beyond Cr, both the absorption sites generate similar moments (Figure~\ref{fig:adatom-MM}). This observed trend in magnetic moment along the 3$d$ series can be understood by simple electron counting. For TM@Hollow cases, two electrons take part in the bonding, and thus do not contribute to moment formation. The rest of the electrons are distributed in the four available $d$-orbitals such that a high-spin solution is favorable.  For example, Mn@Hollow has five electrons that are distributed in four available non-bonding $d$-states, which give rise to a high-spin 3  $\mu_B$ moment. In contrast, four electrons take part in bonding for TM@Valley, which is evident from increased number of TM---P bonds. Rest of the $d$-electrons are distributed in three available $d$-orbitals such that a high-spin solution is favorable. For example, the Mn@Valley has three $d$-electrons that are to be distributed in three non-bonding $d$-orbitals, which consequently generates 3 $\mu_B$  moment (Figure~\ref{fig:adatom-MM}). The low-spin solution is only 30 meV higher in energy in this case. The Fe@Valley is found to be the only exception to this simple model. The model predicts a 2 $\mu_B$ ground state, which we find to be almost 300 meV higher in energy than the non-magnetic ground state. The proximity polarization at the P atoms are found to be very small in these cases ($-$0.04 $\mu_B$ for Cr@Valley to 0.07 $\mu_B$ for Mn@Valley).  Further, these TM@Phosphorene systems must have a semiconducting gap for any plausible spintronic application. Although the intrinsic gap decreases upon TM absorption, we find TM@Valley posses a formidable semiconducting gap in the range 0.4--0.65 eV, except for Sc and Cu (Figure S8 and Table S1 in Supporting Information).  The comparatively smaller gap is due to the appearance of  hybridized $p-d$ states within the gap. It is also interesting to note that the anisotropic dispersion along the zigzag and armchair direction is not affected by the TM adatom absorption, which is intrinsic to single-layer phosphorene.  The phosphorene becomes metallic due to Sc and Cu adatom adsorption (Figure S8) as they dope electron. This is in agreement with the recent experimental observation,~\citep{acs.nanolett.5b03278} and these systems could be used to lower the threshold voltage for $n$-type conduction.

Although TMs bind strongly on pristine phosphorene,  induce local magnetic moment, and also hold semiconducting gap, such TM-phosphorene system may not be a good candidate material for any practical purpose due to very high TM diffusivity. The TM diffusion is an activated process, and on a 2D surface the diffusivity is defined as $D=\frac{1}{4} a_0^2\Gamma$, where $a_0$ is the jump distance, and $\Gamma = \nu \exp(-E_a/k_BT)$ is the Arrhenius jump rate with $\nu$ being the prefactor related to atomic vibration, and $E_a$ is the activation barrier. The activation barrier for valley to valley diffusion is calculated to be 0.69 eV for Co, which correspond to $D_{\rm Co}(300 \rm K)$ = 2.61 $\times$ 10$^{-15}$ cm$^2$/s assuming $\nu=$10$^{13}$ Hz. Thus, due to very high diffusivity, TM induced magnetism on pristine phosphorene is not controllable as the spatial TM distribution strongly depends on time and temperature. In contrast, vacancy defects could act as anchoring site, which substantially increases the activation barrier, and thus TM diffusivity is exponentially reduced. The calculated activation barrier for Co diffusion out of the vacancy defects are calculated to be 1.65 and 1.44 eV for Co@MV and Co@DV complexes, respectively. Such point-defects are present in the as-prepared samples or can be wishfully created on the phosphorene surface. Once absorbed at the defect site, the TM local moment is essentially pinned, and thus the corresponding magnetism will be controllable. The TM@MV have been investigated recently,~\citep{10.1021/jp5110938,10.1021/jp511574n} however, the microscopic picture is not clear yet. Moreover, as we have discussed earlier that the DV is thermodynamically more stable and thus abundant than the MV, which tend to coalesce and form DVs via vacancy diffusion. Therefore, the TM@DV complexes will be particularly more important for any practical purpose toward TM induced spintronics. Here we investigate both TM@MV and TM@DV complexes, and track their electronic and magnetic properties, and provide a microscopic picture. 

{\color{blue}\bf TM-Vacancy complexes.}
The typical atomic configurations of TM@MV and TM@DV are shown in Figures~\ref{fig:structure-defect}(a)-(b). Unlike graphene the TMs are accommodated inside the point-defects due to the intrinsic puckered structure of phosphorene. The TMs are absorbed between the two half-layers [Figure~\ref{fig:structure-defect}(c)], and thus the calculated vertical height $h$ is negative. We find that all metals form two sets of chemical bonds with the under-coordinated P atoms from the defected and undefected half-layers, which we denote as $d_d$ and $d_{ud}$, respectively [Figure~\ref{fig:structure-defect}(a)-(b)].  For TM@MV, metals form two $d_d$ and one $d_{ud}$ bonds. In contrast, two $d_d$ and two $d_{ud}$ bonds are created for TM@DV, and thus the metal atoms move further away from the defected half-layer [Figure~\ref{fig:structure-defect}(c)]. Both $d_d$ and $d_{ud}$ are comparable, and in cases, shorter than the corresponding TM--P covalent bond lengths, and follow similar qualitative trend for TM@MV and TM@DV [Figure~\ref{fig:structure-defect}(d)], which is correlated with the size of the metal ions. The bond lengths first decrease with increasing number of valence electrons for Sc-Co, and then increase for Co-Zn.  It should be noted here that the slight deviation from this general trend for Cr, and Mn [Figure~\ref{fig:structure-defect}(d)] is due to their half-filled $d$ shell, leading to a weaker bonding. 

\begin{figure}[t]
\centering
\includegraphics[scale=0.15]{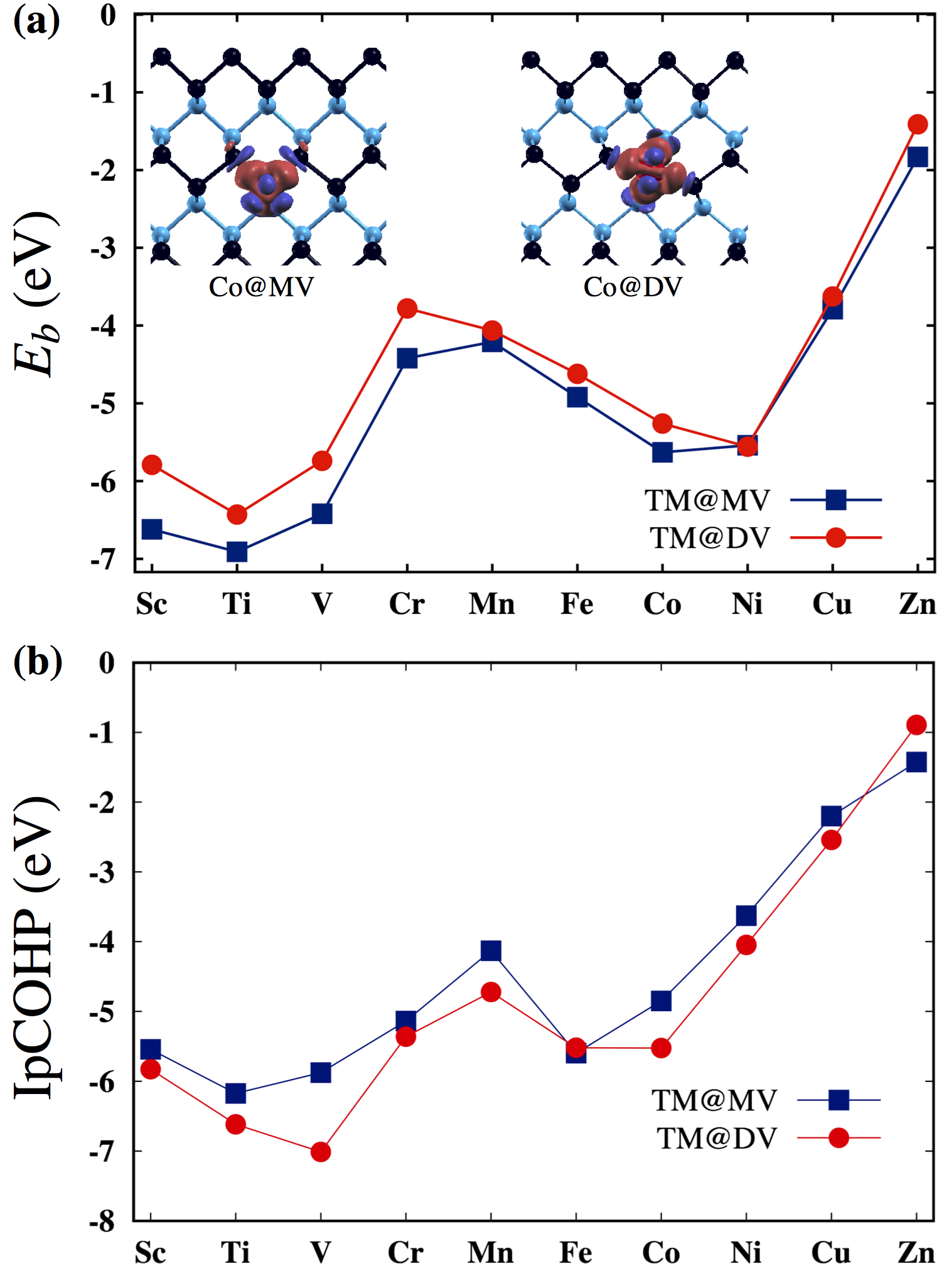}
\caption{(a) Transition metal binding $E_b$ for vacancy-TM complexes.  The TM binding is little stronger in the case of TM@MV compared to TM@DV. The overall trend in $E_b$ across the 3$d$ series is similar to that of other two-dimensional materials.~\citep{PhysRevLett.102.126807,1.4921699}  We argue that this trend could be understood completely by the covalent $p$(P)-$d$(TM) bonding. Representative differential charge densities $\Delta \rho(\mathbf r)$ are shown in the inset for 0.03 $e$/\AA$^3$ isosurface, which indicate strong covalent bonding. The red (blue) color represents electron accumulation (depletion). (b) Calculated integrated partial COHP, which reveal that the qualitative trend in $E_b$ along the 3$d$ series shown in (a) is related to the trend in strong TM--P covalent bonding. }
\label{fig:TM-binding}
\end{figure}

\begin{figure}[t]
\centering
\includegraphics[scale=0.15]{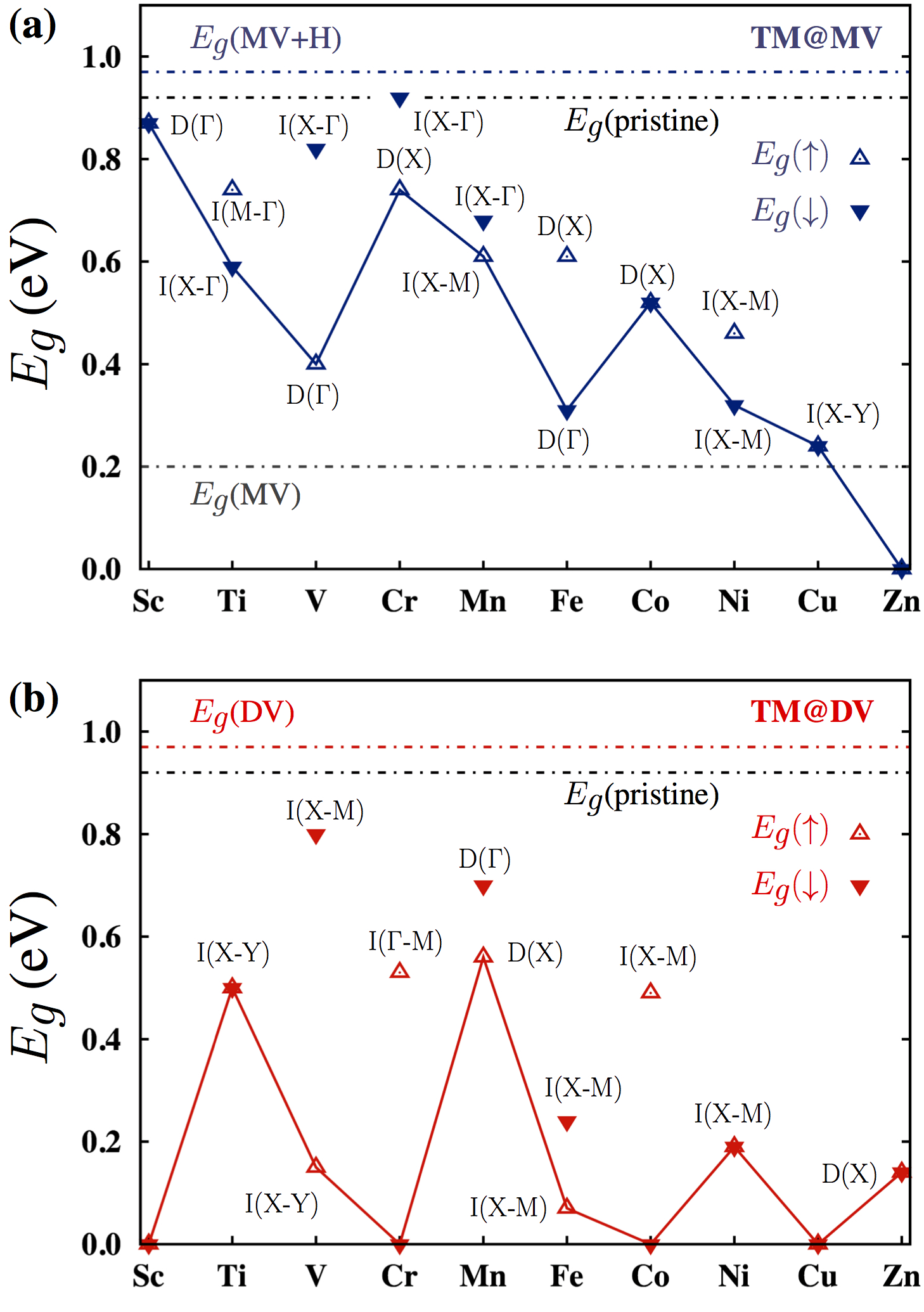}
\caption{Spin-resolved band gaps for (a) TM@MV and (b) TM@DV complexes. The solid line is a guide to the lowest gap along the 3$d$ series.  The nature of the gap is indicated by D (direct) and I (indirect) along with the k-points in the Brillouin zone. I(X-Y) indicates that the VBM is at X-point and CBM is at Y-point in the first Brillouin zone. For TM@MV, the gap decreases as one moves along the 3$d$ series. This trend is nonmonotonous with local structures.  For all TM-doped cases the gap is calculated to be lower than the pristine or defected phosphorene. Moreover, Cr@DV and Co@DV show half-metallic behavior.  The TM@DV complexes (V,  Mn and Fe) show promise as spintronic materials with reasonable band gaps.}
\label{fig:gap}
\end{figure}


We find most of the metals bind very strongly with the point defects [Figure~\ref{fig:TM-binding}(a)] with binding energy as strong (-7 to -4 eV) as in graphene, and also follow similar non-monotonic behavior.~\citep{PhysRevLett.102.126807} To analyze the nature of electronic interaction between the defected phosphorene and the absorbed metal, we calculated the differential charge density $\Delta \rho(\mathbf r)$, which is shown in the inset of Figure~\ref{fig:TM-binding}(a). The calculated $\Delta \rho(\mathbf r)$ is typical of covalent bonding, which is reflected in strong TM--P bonding.  This covalent bonding is also emulated in the respective TM--P bond lengths (Figure~\ref{fig:structure-defect}). The observed trend in binding can be explained by carefully looking into the TM--P bonding. The bare MV (DV) consists three (four) dangling bonds, which are completely saturated by TM incorporation. The TM-defect binding is mainly derived by the hybridization between TM-3$d$ and 3$p$ orbitals of the under-coordinated  P atoms. We find that the three bonding states are derived from P-3$p$ and TM-$d_{xy}$/$d_{z^2}$ orbitals for TM@MV. In contrast, the four bonding states for TM@DV are derived from TM-$d_{x^2-y^2}$/$d_{yz}$ and the neighbouring P-$p$ orbitals.

The TM binding/absorption energy decreases for Ti-Cr, which in contrast increases as we move along Cr to Ni [Figure~\ref{fig:TM-binding}(a)].  The binding for Cu and Zn is comparatively much weaker due to their filled atomic orbitals. 
Such evolution in binding along the 3$d$ series is generic to TM absorption on 2D materials but it still lacks a microscopic description.~\citep{PhysRevLett.102.126807,1.4921699} To corroborate the strong TM binding and the overall trend in Figure~\ref{fig:TM-binding}(a), we have calculated the TM--P projected COHP (Figure S9 in Supplementary Information), which is the density of states weighted by the corresponding Hamiltonian matrix element.~\citep{doi:10.1021/jp202489s, doi:10.1021/j100135a014}  The COHP is an excellent tool to analyze the strength and nature of bonding.   The bonding states are dominant in the valence band region indicating strong covalent TM--P interactions (Figure S9), which lead to strong metal binding. In order to quantify the TM--P covalent interaction, we have integrated the COHP up to the Fermi level, and summed over all the TM--P bonds [IpCOHP in Figure~\ref{fig:TM-binding}(b)].  The calculated IpCOHP and binding energy show similar qualitative evolution along the 3$d$ series. This indicates that the differential covalent bonding is responsible for the observed trend in binding energy.  For example,  the Ti--P covalent bonding [IpCOHP$_{\rm Ti-P}$ = $-$6.61 eV] is much stronger than the Cr--P bonding [IpCOHP$_{\rm Cr-P}$ = $-$5.36 eV], which is reflected in their respective binding energies for DV complexes.

Now it would be interesting to investigate the effect of TM absorption at the defect site on the intrinsic semiconducting nature of phosphorene. The defect free phosphorene is a direct gap semiconductor with 0.9 eV gap, which is found to be robust against the formation of point-defects as we have discussed earlier. Calculated spin-resolved band gaps are shown in Figure~\ref{fig:gap} for both TM@MV and TM@DV complexes (Table S2 and S3 in Supporting Information). The in-gap states appear due to metal absorption, which result in band gaps that are smaller compared to the pristine and defected bare phosphorene. Further, the gap decreases along the 3$d$ series for TM@MV complexes, however we observe this decrease to be nonmonotonus with local peaks for (Cr/Mn/Co)@MV. In general, we find that CBM originates from TM-$d$ orbitals, whereas the VBM is induced by either P-$p$ or non-bonding TM-$d$ orbitals. In contrast, while both VBM and CBM originate from TM-$d$ orbitals, the calculated gaps are found to be much smaller.      

In this regard, the case of metals absorbed at the divacancy is particularly interesting. Although, the band gaps are further reduced, the Ti, V, Mn, Fe, and Ni complexes retain gaps ranging between 0.1$-$0.6 eV.  Reminding that the conventional generalized gradient approximation to the exchange-correlation functional underestimates the band gap, the true gaps should be larger than the PBE gaps reported in Figure~\ref{fig:gap}(b). Indeed while the PBE functional predicts a smaller gap of 0.1 eV for Fe@DV, the hybrid HSE06 functional estimates a much larger gap of 1.01 eV (Figure S10), which may be sufficient for spintronic devices.  Moreover, metal doping at the divacancy could completely alter the intrinsic band structure from semiconducting phosphorene to half-metallic (Cr/Co)@DV or metallic (Sc/Cu)@DV complexes.

\begin{figure}[t]
\centering
\includegraphics[scale=0.16]{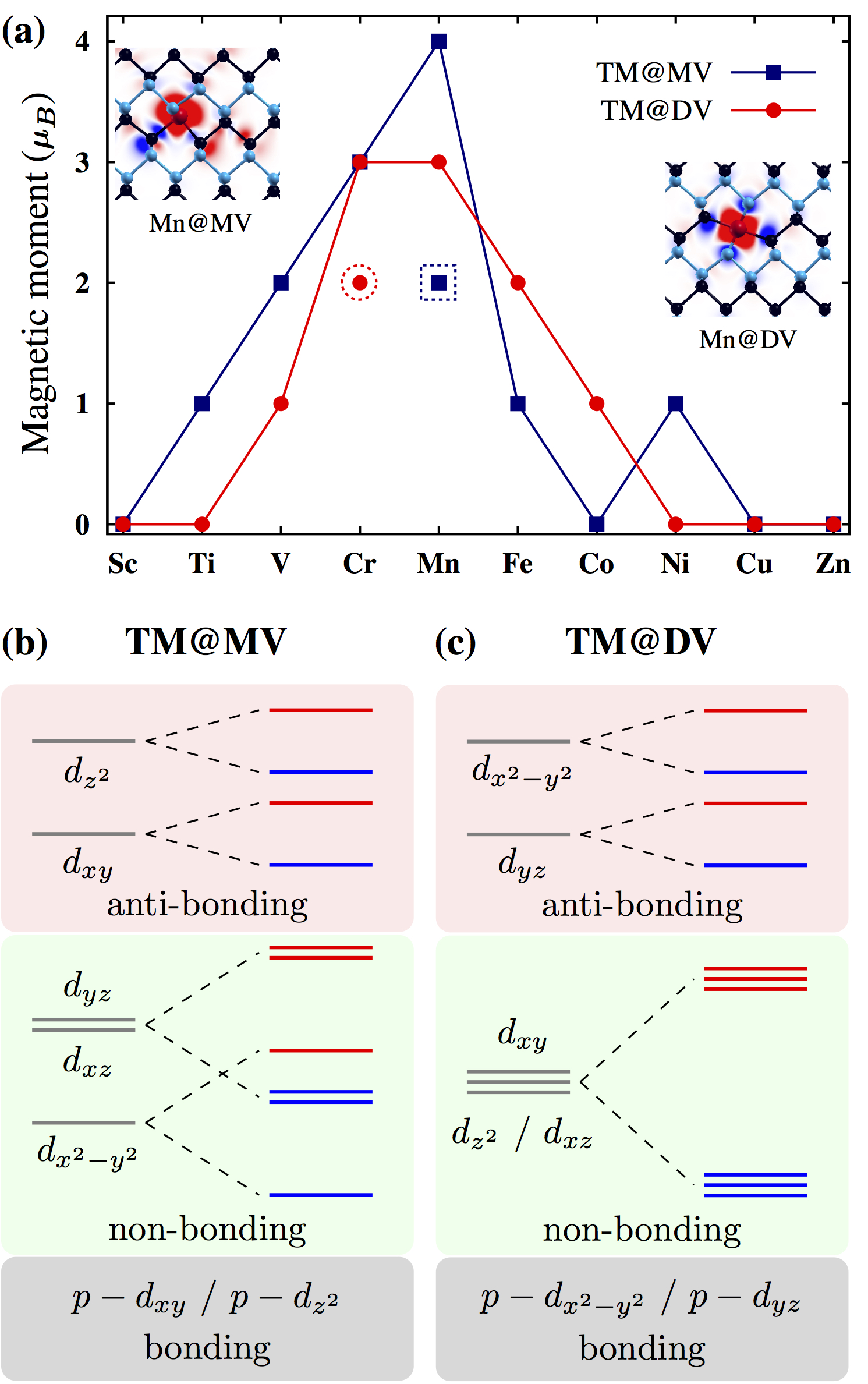}
\caption{(a) Calculated magnetic moments of TM-vacancy complexes, with increasing valence electrons. For Mn@MV and Cr@DV, the solutions with 2$\mu_B$ moments are very close in energy with their respective ground states (within 60 meV), which are also highlighted. Representative magnetization densities are shown in the inset. (b)-(c) Simple model to describe the observed trend in magnetic moment in (a) for TM@MV and TM@DV, respectively. In both the cases, the electrons are distributed in the non-bonding states such that it gives rise to high-spin solutions.}
\label{fig:defect-moment}
\end{figure}

Now we turn our attention to the local moment created due to metal absorption at the point-defects. The calculated moments are shown in Figure~\ref{fig:defect-moment}(a), which show a dome-shaped behavior for both TM@MV and TM@DV as the number of $d$-electrons increases. The density of states and magnetization density analyses [Figure~\ref{fig:defect-moment}(a)] indicate that the moment originates from the localized nonbonding $d$-orbitals, which also polarize the neighbouring P atoms. However, the proximity polarization of the P atoms are found be small,  $-$0.09 (Cr@MV) to 0.02 $\mu_B$ (Mn@MV) for TMs absorbed at the monovacancy, and $-$0.07 (Mn@DV) to 0.07 $\mu_B$ (Fe@DV) for TMs absorbed at the divacancy.  The local moment at Ni (0.50 $\mu_B$) absorbed at the MV is found to be fragile with small exchange splitting and spin-flip energy.   

By comparing the band structure and density of states of the TM-vacancy complexes to those with the defected phosphorene, we developed a simple model to explain the trends in magnetic moment [Figure~\ref{fig:defect-moment}(b)-(c)].  Removal of a single atom from phosphorene creates three dangling bonds, and thus, while a TM atom is placed at the MV, three of TM valence electrons form covalent TM--P bonds. We find these to be two $d_{xy}-p(\rm P)$ and one  $d_{z^2}-p(\rm P)$ covalent bonds [Figure~\ref{fig:defect-moment}(b)]. For example, three valence electrons of Sc completely saturates three MV dangling bonds, and thus Sc@MV does not produce any local moment.  Next, for Ti-Co, rest of the TM electrons occupy the non-bonding $d$ states, $d_{x^2-y^2}$, $d_{xz}$, and $d_{yz}$ [Figure~\ref{fig:defect-moment}(b)], such that it always produces a high-spin solution. For example, Cr has six valence electrons, which give rise to 3 $\mu_B$ moment. Once these non-bonding states are filled, the TM electrons occupy the anti-bonding $d_{xy}$ and $d_{z^2}$ states for Ni--Zn.  The Mn@MV is found to be the only exception which does not follow this simple rule. However, the predicted 4 $\mu_B$ solution is only 60 meV lower in energy than the 2 $\mu_B$ solution, which is in accordance with the proposed model.

In contrast, four additional electrons are required to completely saturate the four dangling bonds in unrelaxed divacancy. Thus, for TM@DV, the TMs form four strong covalent bonds with the dangling P atoms. 
We find these to be two $d_{x^2-y^2}-p(\rm P)$ and two  $d_{yz}-p(\rm P)$ covalent bonds. Thus, all the four valence electrons of Ti are consumed in TM--P covalent bonding, and Ti does not produce any local moment. For V--Ni, the rest of the $d$ electrons occupy the non-bonding $d_{xy}$, $d_{xz}$ and $d_{z^2}$ orbitals [Figure~\ref{fig:defect-moment}(c)], such that it produces a high-spin solution. Next, the electrons in Cu and Zn are occupied in the anti-bonding $d_{yz}$ and $d_{x^2-y^2}$ orbitals. The Cr@DV complex is the only exception to this model, which predicts a 2 $\mu_B$ ground state. Instead, we find the 3 $\mu_B$ solution to be 40 meV lower in energy.

The magnetic exchange splitting $\delta E_{ex}$ between the majority and minority spin bands is the key to the magnetism. The calculated $\delta E_{ex}$ for the systems with non-zero local moments (Figure~\ref{fig:defect-moment}) are found to be quite large 0.40--2.50 eV (Table S4 in Supplementary Information), and is linearly correlated with the local moment. The spin imbalance owing to the exchange splitting produces the magnetic moment, and therefore the moment increases with exchange splitting. Further, the calculated exchange splitting is found to be larger in DV@TM than for MV@TM. Such large $\delta E_{ex}$ is comparable to the corresponding magnetic TM bulk, and ultrathin TM films on nonmagnetic substrate.~\citep{doi:10.1080, PhysRevLett.67.2363, PhysRevB.47.16441, PhysRevB.50.17496} For example, the calculated $\delta E_{ex}$ for Fe@DV (2.48 eV) is substantially larger than epitaxial ML-Fe/Cu(100) [0.9--1.2 eV] and ML-Fe/Ag(100) [1.8 eV] that are measured via inverse photoelectron spectroscopy.~\citep{PhysRevLett.67.2363} There could be some variation in $\delta E_{ex}$ with momentum ${\bm {k}}$, however, we do not observe any noticeable momentum variation due to dispersion less non-bonding $d$ states. Another important quantity is the energy required for spin-flip transitions, which can either increase or decrease the local moment. We have calculated the moment diminishing spin-flip energies $\delta E_{sf}$, which is related to the survival of the local moment. Similar to exchange splitting, the calculated $\delta E_{sf}$ are also found to be quite large (Table S4 in Supplementary Information).

{\color{blue}\bf Summary.} 
We investigate the electronic and magnetic properties of transition-metals absorbed on the pristine and defected single-layer phosphorene. Inducing local moment in otherwise non-magnetic semiconductor is the key first step toward plausible spintronic application. It was earlier believed that TMs are absorbed at the hollow site, however, we find that TMs favor the valley site, which is in between the two half-layers. Metals absorbed at this new site have completely different electronic and magnetic properties. Metals bind strongly on the pristine phosphorene, induce a local moment, and the semiconducting gap is retained. However, a very high metal diffusivity on the pristine phosphorene make them unfavorable for any practical purpose. Their high mobility leading to time-evolving spatial TM distribution makes the induced magnetism uncontrollable. Thus, one require the metals to be absorbed at the point-defects, which are present in as-prepared phosphorene or could be wishfully created. Moreover, the intrinsic semiconducting gap is found to be robust against the formation of such point-defects. These vacancy defects act as anchoring sites, where the metal atoms bind very strongly with phosphorene through strong covalent TM--P interaction. The evolution of metal absorption energy along the 3$d$ series follow a generic trend as observed for other 2D materials, which we have qualitatively explained via detailed COHP calculations. 

We propose that due to much higher thermodynamic stability of the bare divacancy defects over monovacancy, the TM@DV complexes are better proposition for spintronic applications. Although the calculated gaps for TM@DV complexes are relatively lower than the gaps for pristine and defected bare phosphorene, some of these complexes retain a formidable gap (1.01 eV for Fe@DV) along with a local moment that may be useful for spintronic application. The 3$d$ evolution in local moment for TM-defect complexes has been explained within a simple model. Moreover, the magnetism in these TM-defect complexes are stable with large exchange splitting and spin-flip energy together. Thus, we argue that V, Mn, and Fe absorbed at the divacancy defect to be good candidates for spintronic device applications. We also note that metal absorption could completely alter the intrinsic semiconducting nature of the single-layer phosphorene and give rise to half-metallic [(Cr/Co)@DV] or metallic [(Sc/Cu)@DV] composite system. Although the present study indicate the TM@DV complexes to be encouraging candidates for spintronic device application, further theoretical and experimental efforts are necessary concerning long-range magnetic ordering, and corresponding transition temperature. We hope the present study will encourage such investigations in future. 

\begin{acknowledgements}
MK acknowledges a grant from the Department of Science and Technology, India under the Ramanujan Fellowship. The authors acknowledge the supercomputing facilities at the Centre for Development of Advanced Computing, Pune; Inter University Accelerator Centre, Delhi; and at the Center for Computational Materials Science, Institute of Materials Research, Tohoku University. 
\end{acknowledgements}


\begin{thebibliography}{70}%
\makeatletter
\providecommand \@ifxundefined [1]{%
 \@ifx{#1\undefined}
}%
\providecommand \@ifnum [1]{%
 \ifnum #1\expandafter \@firstoftwo
 \else \expandafter \@secondoftwo
 \fi
}%
\providecommand \@ifx [1]{%
 \ifx #1\expandafter \@firstoftwo
 \else \expandafter \@secondoftwo
 \fi
}%
\providecommand \natexlab [1]{#1}%
\providecommand \enquote  [1]{``#1''}%
\providecommand \bibnamefont  [1]{#1}%
\providecommand \bibfnamefont [1]{#1}%
\providecommand \citenamefont [1]{#1}%
\providecommand \href@noop [0]{\@secondoftwo}%
\providecommand \href [0]{\begingroup \@sanitize@url \@href}%
\providecommand \@href[1]{\@@startlink{#1}\@@href}%
\providecommand \@@href[1]{\endgroup#1\@@endlink}%
\providecommand \@sanitize@url [0]{\catcode `\\12\catcode `\$12\catcode
  `\&12\catcode `\#12\catcode `\^12\catcode `\_12\catcode `\%12\relax}%
\providecommand \@@startlink[1]{}%
\providecommand \@@endlink[0]{}%
\providecommand \url  [0]{\begingroup\@sanitize@url \@url }%
\providecommand \@url [1]{\endgroup\@href {#1}{\urlprefix }}%
\providecommand \urlprefix  [0]{URL }%
\providecommand \Eprint [0]{\href }%
\providecommand \doibase [0]{http://dx.doi.org/}%
\providecommand \selectlanguage [0]{\@gobble}%
\providecommand \bibinfo  [0]{\@secondoftwo}%
\providecommand \bibfield  [0]{\@secondoftwo}%
\providecommand \translation [1]{[#1]}%
\providecommand \BibitemOpen [0]{}%
\providecommand \bibitemStop [0]{}%
\providecommand \bibitemNoStop [0]{.\EOS\space}%
\providecommand \EOS [0]{\spacefactor3000\relax}%
\providecommand \BibitemShut  [1]{\csname bibitem#1\endcsname}%
\let\auto@bib@innerbib\@empty
\bibitem [{\citenamefont {Geim}\ and\ \citenamefont
  {Novoselov}(2007)}]{10.1038/nmat1849}%
  \BibitemOpen
  \bibfield  {author} {\bibinfo {author} {\bibfnamefont {A.~K.}\ \bibnamefont
  {Geim}}\ and\ \bibinfo {author} {\bibfnamefont {K.~S.}\ \bibnamefont
  {Novoselov}},\ }\href@noop {} {\bibfield  {journal} {\bibinfo  {journal}
  {Nat. Mater.}\ }\textbf {\bibinfo {volume} {6}},\ \bibinfo {pages} {183}
  (\bibinfo {year} {2007})}\BibitemShut {NoStop}%
\bibitem [{\citenamefont {Schwierz}(2010)}]{10.1038/nnano.2010.89}%
  \BibitemOpen
  \bibfield  {author} {\bibinfo {author} {\bibfnamefont {F.}~\bibnamefont
  {Schwierz}},\ }\href@noop {} {\bibfield  {journal} {\bibinfo  {journal} {Nat.
  Nano.}\ }\textbf {\bibinfo {volume} {5}},\ \bibinfo {pages} {487} (\bibinfo
  {year} {2010})}\BibitemShut {NoStop}%
\bibitem [{\citenamefont {Novoselov}\ \emph {et~al.}(2012)\citenamefont
  {Novoselov}, \citenamefont {Falko}, \citenamefont {Colombo}, \citenamefont
  {Gellert}, \citenamefont {Schwab},\ and\ \citenamefont
  {Kim}}]{10.1038/nature11458}%
  \BibitemOpen
  \bibfield  {author} {\bibinfo {author} {\bibfnamefont {K.~S.}\ \bibnamefont
  {Novoselov}}, \bibinfo {author} {\bibfnamefont {V.~I.}\ \bibnamefont
  {Falko}}, \bibinfo {author} {\bibfnamefont {L.}~\bibnamefont {Colombo}},
  \bibinfo {author} {\bibfnamefont {P.~R.}\ \bibnamefont {Gellert}}, \bibinfo
  {author} {\bibfnamefont {M.~G.}\ \bibnamefont {Schwab}}, \ and\ \bibinfo
  {author} {\bibfnamefont {K.}~\bibnamefont {Kim}},\ }\href@noop {} {\bibfield
  {journal} {\bibinfo  {journal} {Nature}\ }\textbf {\bibinfo {volume} {490}},\
  \bibinfo {pages} {192} (\bibinfo {year} {2012})}\BibitemShut {NoStop}%
\bibitem [{\citenamefont {Butler}\ \emph {et~al.}(2013)\citenamefont {Butler},
  \citenamefont {Hollen}, \citenamefont {Cao}, \citenamefont {Cui},
  \citenamefont {Gupta}, \citenamefont {Gutiérrez}, \citenamefont {Heinz},
  \citenamefont {Hong}, \citenamefont {Huang}, \citenamefont {Ismach},
  \citenamefont {Johnston-Halperin}, \citenamefont {Kuno}, \citenamefont
  {Plashnitsa}, \citenamefont {Robinson}, \citenamefont {Ruoff}, \citenamefont
  {Salahuddin}, \citenamefont {Shan}, \citenamefont {Shi}, \citenamefont
  {Spencer}, \citenamefont {Terrones}, \citenamefont {Windl},\ and\
  \citenamefont {Goldberger}}]{10.1021/nn400280c}%
  \BibitemOpen
  \bibfield  {author} {\bibinfo {author} {\bibfnamefont {S.~Z.}\ \bibnamefont
  {Butler}}, \bibinfo {author} {\bibfnamefont {S.~M.}\ \bibnamefont {Hollen}},
  \bibinfo {author} {\bibfnamefont {L.}~\bibnamefont {Cao}}, \bibinfo {author}
  {\bibfnamefont {Y.}~\bibnamefont {Cui}}, \bibinfo {author} {\bibfnamefont
  {J.~A.}\ \bibnamefont {Gupta}}, \bibinfo {author} {\bibfnamefont {H.~R.}\
  \bibnamefont {Gutiérrez}}, \bibinfo {author} {\bibfnamefont {T.~F.}\
  \bibnamefont {Heinz}}, \bibinfo {author} {\bibfnamefont {S.~S.}\ \bibnamefont
  {Hong}}, \bibinfo {author} {\bibfnamefont {J.}~\bibnamefont {Huang}},
  \bibinfo {author} {\bibfnamefont {A.~F.}\ \bibnamefont {Ismach}}, \bibinfo
  {author} {\bibfnamefont {E.}~\bibnamefont {Johnston-Halperin}}, \bibinfo
  {author} {\bibfnamefont {M.}~\bibnamefont {Kuno}}, \bibinfo {author}
  {\bibfnamefont {V.~V.}\ \bibnamefont {Plashnitsa}}, \bibinfo {author}
  {\bibfnamefont {R.~D.}\ \bibnamefont {Robinson}}, \bibinfo {author}
  {\bibfnamefont {R.~S.}\ \bibnamefont {Ruoff}}, \bibinfo {author}
  {\bibfnamefont {S.}~\bibnamefont {Salahuddin}}, \bibinfo {author}
  {\bibfnamefont {J.}~\bibnamefont {Shan}}, \bibinfo {author} {\bibfnamefont
  {L.}~\bibnamefont {Shi}}, \bibinfo {author} {\bibfnamefont {M.~G.}\
  \bibnamefont {Spencer}}, \bibinfo {author} {\bibfnamefont {M.}~\bibnamefont
  {Terrones}}, \bibinfo {author} {\bibfnamefont {W.}~\bibnamefont {Windl}}, \
  and\ \bibinfo {author} {\bibfnamefont {J.~E.}\ \bibnamefont {Goldberger}},\
  }\href@noop {} {\bibfield  {journal} {\bibinfo  {journal} {ACS Nano}\
  }\textbf {\bibinfo {volume} {7}},\ \bibinfo {pages} {2898} (\bibinfo {year}
  {2013})}\BibitemShut {NoStop}%
\bibitem [{\citenamefont {Morozov}\ \emph {et~al.}(2008)\citenamefont
  {Morozov}, \citenamefont {Novoselov}, \citenamefont {Katsnelson},
  \citenamefont {Schedin}, \citenamefont {Elias}, \citenamefont {Jaszczak},\
  and\ \citenamefont {Geim}}]{PhysRevLett.100.016602}%
  \BibitemOpen
  \bibfield  {author} {\bibinfo {author} {\bibfnamefont {S.~V.}\ \bibnamefont
  {Morozov}}, \bibinfo {author} {\bibfnamefont {K.~S.}\ \bibnamefont
  {Novoselov}}, \bibinfo {author} {\bibfnamefont {M.~I.}\ \bibnamefont
  {Katsnelson}}, \bibinfo {author} {\bibfnamefont {F.}~\bibnamefont {Schedin}},
  \bibinfo {author} {\bibfnamefont {D.~C.}\ \bibnamefont {Elias}}, \bibinfo
  {author} {\bibfnamefont {J.~A.}\ \bibnamefont {Jaszczak}}, \ and\ \bibinfo
  {author} {\bibfnamefont {A.~K.}\ \bibnamefont {Geim}},\ }\href@noop {}
  {\bibfield  {journal} {\bibinfo  {journal} {Phys. Rev. Lett.}\ }\textbf
  {\bibinfo {volume} {100}},\ \bibinfo {pages} {016602} (\bibinfo {year}
  {2008})}\BibitemShut {NoStop}%
\bibitem [{\citenamefont {Wallace}(1947)}]{PhysRev.71.622}%
  \BibitemOpen
  \bibfield  {author} {\bibinfo {author} {\bibfnamefont {P.~R.}\ \bibnamefont
  {Wallace}},\ }\href@noop {} {\bibfield  {journal} {\bibinfo  {journal} {Phys.
  Rev.}\ }\textbf {\bibinfo {volume} {71}},\ \bibinfo {pages} {622} (\bibinfo
  {year} {1947})}\BibitemShut {NoStop}%
\bibitem [{\citenamefont {Son}\ \emph {et~al.}(2006)\citenamefont {Son},
  \citenamefont {Cohen},\ and\ \citenamefont {Louie}}]{PhysRevLett.97.216803}%
  \BibitemOpen
  \bibfield  {author} {\bibinfo {author} {\bibfnamefont {Y.-W.}\ \bibnamefont
  {Son}}, \bibinfo {author} {\bibfnamefont {M.~L.}\ \bibnamefont {Cohen}}, \
  and\ \bibinfo {author} {\bibfnamefont {S.~G.}\ \bibnamefont {Louie}},\
  }\href@noop {} {\bibfield  {journal} {\bibinfo  {journal} {Phys. Rev. Lett.}\
  }\textbf {\bibinfo {volume} {97}},\ \bibinfo {pages} {216803} (\bibinfo
  {year} {2006})}\BibitemShut {NoStop}%
\bibitem [{\citenamefont {Zhou}\ \emph {et~al.}(2007)\citenamefont {Zhou},
  \citenamefont {Gweon}, \citenamefont {Fedorov}, \citenamefont {First},
  \citenamefont {de~Heer}, \citenamefont {Lee}, \citenamefont {Guinea},
  \citenamefont {Castro~Neto},\ and\ \citenamefont
  {Lanzara}}]{10.1038/nmat2003}%
  \BibitemOpen
  \bibfield  {author} {\bibinfo {author} {\bibfnamefont {S.~Y.}\ \bibnamefont
  {Zhou}}, \bibinfo {author} {\bibfnamefont {G.~H.}\ \bibnamefont {Gweon}},
  \bibinfo {author} {\bibfnamefont {A.~V.}\ \bibnamefont {Fedorov}}, \bibinfo
  {author} {\bibfnamefont {P.~N.}\ \bibnamefont {First}}, \bibinfo {author}
  {\bibfnamefont {W.~A.}\ \bibnamefont {de~Heer}}, \bibinfo {author}
  {\bibfnamefont {D.~H.}\ \bibnamefont {Lee}}, \bibinfo {author} {\bibfnamefont
  {F.}~\bibnamefont {Guinea}}, \bibinfo {author} {\bibfnamefont {A.~H.}\
  \bibnamefont {Castro~Neto}}, \ and\ \bibinfo {author} {\bibfnamefont
  {A.}~\bibnamefont {Lanzara}},\ }\href@noop {} {\bibfield  {journal} {\bibinfo
   {journal} {Nat. Mater.}\ }\textbf {\bibinfo {volume} {6}},\ \bibinfo {pages}
  {770} (\bibinfo {year} {2007})}\BibitemShut {NoStop}%
\bibitem [{\citenamefont {Balog}\ \emph {et~al.}(2010)\citenamefont {Balog},
  \citenamefont {Jorgensen}, \citenamefont {Nilsson}, \citenamefont {Andersen},
  \citenamefont {Rienks}, \citenamefont {Bianchi}, \citenamefont {Fanetti},
  \citenamefont {Laegsgaard}, \citenamefont {Baraldi}, \citenamefont {Lizzit},
  \citenamefont {Sljivancanin}, \citenamefont {Besenbacher}, \citenamefont
  {Hammer}, \citenamefont {Pedersen}, \citenamefont {Hofmann},\ and\
  \citenamefont {Hornekaer}}]{10.1038/nmat2710}%
  \BibitemOpen
  \bibfield  {author} {\bibinfo {author} {\bibfnamefont {R.}~\bibnamefont
  {Balog}}, \bibinfo {author} {\bibfnamefont {B.}~\bibnamefont {Jorgensen}},
  \bibinfo {author} {\bibfnamefont {L.}~\bibnamefont {Nilsson}}, \bibinfo
  {author} {\bibfnamefont {M.}~\bibnamefont {Andersen}}, \bibinfo {author}
  {\bibfnamefont {E.}~\bibnamefont {Rienks}}, \bibinfo {author} {\bibfnamefont
  {M.}~\bibnamefont {Bianchi}}, \bibinfo {author} {\bibfnamefont
  {M.}~\bibnamefont {Fanetti}}, \bibinfo {author} {\bibfnamefont
  {E.}~\bibnamefont {Laegsgaard}}, \bibinfo {author} {\bibfnamefont
  {A.}~\bibnamefont {Baraldi}}, \bibinfo {author} {\bibfnamefont
  {S.}~\bibnamefont {Lizzit}}, \bibinfo {author} {\bibfnamefont
  {Z.}~\bibnamefont {Sljivancanin}}, \bibinfo {author} {\bibfnamefont
  {F.}~\bibnamefont {Besenbacher}}, \bibinfo {author} {\bibfnamefont
  {B.}~\bibnamefont {Hammer}}, \bibinfo {author} {\bibfnamefont {T.~G.}\
  \bibnamefont {Pedersen}}, \bibinfo {author} {\bibfnamefont {P.}~\bibnamefont
  {Hofmann}}, \ and\ \bibinfo {author} {\bibfnamefont {L.}~\bibnamefont
  {Hornekaer}},\ }\href@noop {} {\bibfield  {journal} {\bibinfo  {journal}
  {Nat. Mater.}\ }\textbf {\bibinfo {volume} {9}},\ \bibinfo {pages} {315}
  (\bibinfo {year} {2010})}\BibitemShut {NoStop}%
\bibitem [{\citenamefont {Mak}\ \emph {et~al.}(2009)\citenamefont {Mak},
  \citenamefont {Lui}, \citenamefont {Shan},\ and\ \citenamefont
  {Heinz}}]{PhysRevLett.102.256405}%
  \BibitemOpen
  \bibfield  {author} {\bibinfo {author} {\bibfnamefont {K.~F.}\ \bibnamefont
  {Mak}}, \bibinfo {author} {\bibfnamefont {C.~H.}\ \bibnamefont {Lui}},
  \bibinfo {author} {\bibfnamefont {J.}~\bibnamefont {Shan}}, \ and\ \bibinfo
  {author} {\bibfnamefont {T.~F.}\ \bibnamefont {Heinz}},\ }\href@noop {}
  {\bibfield  {journal} {\bibinfo  {journal} {Phys. Rev. Lett.}\ }\textbf
  {\bibinfo {volume} {102}},\ \bibinfo {pages} {256405} (\bibinfo {year}
  {2009})}\BibitemShut {NoStop}%
\bibitem [{\citenamefont {Choi}\ \emph {et~al.}(2010)\citenamefont {Choi},
  \citenamefont {Jhi},\ and\ \citenamefont {Son}}]{PhysRevB.81.081407}%
  \BibitemOpen
  \bibfield  {author} {\bibinfo {author} {\bibfnamefont {S.-M.}\ \bibnamefont
  {Choi}}, \bibinfo {author} {\bibfnamefont {S.-H.}\ \bibnamefont {Jhi}}, \
  and\ \bibinfo {author} {\bibfnamefont {Y.-W.}\ \bibnamefont {Son}},\
  }\href@noop {} {\bibfield  {journal} {\bibinfo  {journal} {Phys. Rev. B}\
  }\textbf {\bibinfo {volume} {81}},\ \bibinfo {pages} {081407} (\bibinfo
  {year} {2010})}\BibitemShut {NoStop}%
\bibitem [{\citenamefont {Wang}\ \emph
  {et~al.}(2012{\natexlab{a}})\citenamefont {Wang}, \citenamefont
  {Kalantar-Zadeh}, \citenamefont {Kis}, \citenamefont {Coleman},\ and\
  \citenamefont {Strano}}]{10.1038/nnano.2012.193}%
  \BibitemOpen
  \bibfield  {author} {\bibinfo {author} {\bibfnamefont {Q.~H.}\ \bibnamefont
  {Wang}}, \bibinfo {author} {\bibfnamefont {K.}~\bibnamefont
  {Kalantar-Zadeh}}, \bibinfo {author} {\bibfnamefont {A.}~\bibnamefont {Kis}},
  \bibinfo {author} {\bibfnamefont {J.~N.}\ \bibnamefont {Coleman}}, \ and\
  \bibinfo {author} {\bibfnamefont {M.~S.}\ \bibnamefont {Strano}},\
  }\href@noop {} {\bibfield  {journal} {\bibinfo  {journal} {Nat. Nano.}\
  }\textbf {\bibinfo {volume} {7}},\ \bibinfo {pages} {699} (\bibinfo {year}
  {2012}{\natexlab{a}})}\BibitemShut {NoStop}%
\bibitem [{\citenamefont {Osada}\ and\ \citenamefont
  {Sasaki}(2012)}]{ADMA201103241}%
  \BibitemOpen
  \bibfield  {author} {\bibinfo {author} {\bibfnamefont {M.}~\bibnamefont
  {Osada}}\ and\ \bibinfo {author} {\bibfnamefont {T.}~\bibnamefont {Sasaki}},\
  }\href@noop {} {\bibfield  {journal} {\bibinfo  {journal} {Adv. Mater.}\
  }\textbf {\bibinfo {volume} {24}},\ \bibinfo {pages} {210} (\bibinfo {year}
  {2012})}\BibitemShut {NoStop}%
\bibitem [{\citenamefont {Late}\ \emph {et~al.}(2012)\citenamefont {Late},
  \citenamefont {Liu}, \citenamefont {Luo}, \citenamefont {Yan}, \citenamefont
  {Matte}, \citenamefont {Grayson}, \citenamefont {Rao},\ and\ \citenamefont
  {Dravid}}]{ADMA201201361}%
  \BibitemOpen
  \bibfield  {author} {\bibinfo {author} {\bibfnamefont {D.~J.}\ \bibnamefont
  {Late}}, \bibinfo {author} {\bibfnamefont {B.}~\bibnamefont {Liu}}, \bibinfo
  {author} {\bibfnamefont {J.}~\bibnamefont {Luo}}, \bibinfo {author}
  {\bibfnamefont {A.}~\bibnamefont {Yan}}, \bibinfo {author} {\bibfnamefont
  {H.~S. S.~R.}\ \bibnamefont {Matte}}, \bibinfo {author} {\bibfnamefont
  {M.}~\bibnamefont {Grayson}}, \bibinfo {author} {\bibfnamefont {C.~N.~R.}\
  \bibnamefont {Rao}}, \ and\ \bibinfo {author} {\bibfnamefont {V.~P.}\
  \bibnamefont {Dravid}},\ }\href@noop {} {\bibfield  {journal} {\bibinfo
  {journal} {Adv. Mater.}\ }\textbf {\bibinfo {volume} {24}},\ \bibinfo {pages}
  {3549} (\bibinfo {year} {2012})}\BibitemShut {NoStop}%
\bibitem [{\citenamefont {Radisavljevic}\ \emph {et~al.}(2011)\citenamefont
  {Radisavljevic}, \citenamefont {Radenovic}, \citenamefont {Brivio},
  \citenamefont {Giacometti},\ and\ \citenamefont
  {Kis}}]{10.1038/nnano.2010.279}%
  \BibitemOpen
  \bibfield  {author} {\bibinfo {author} {\bibfnamefont {B.}~\bibnamefont
  {Radisavljevic}}, \bibinfo {author} {\bibfnamefont {A.}~\bibnamefont
  {Radenovic}}, \bibinfo {author} {\bibfnamefont {J.}~\bibnamefont {Brivio}},
  \bibinfo {author} {\bibfnamefont {V.}~\bibnamefont {Giacometti}}, \ and\
  \bibinfo {author} {\bibfnamefont {A.}~\bibnamefont {Kis}},\ }\href@noop {}
  {\bibfield  {journal} {\bibinfo  {journal} {Nat. Nano.}\ }\textbf {\bibinfo
  {volume} {6}},\ \bibinfo {pages} {147} (\bibinfo {year} {2011})}\BibitemShut
  {NoStop}%
\bibitem [{\citenamefont {Li}\ \emph {et~al.}(2014)\citenamefont {Li},
  \citenamefont {Yu}, \citenamefont {Ye}, \citenamefont {Ge}, \citenamefont
  {Ou}, \citenamefont {Wu}, \citenamefont {Feng}, \citenamefont {Chen},\ and\
  \citenamefont {Zhang}}]{10.1038/nnano.2014.35}%
  \BibitemOpen
  \bibfield  {author} {\bibinfo {author} {\bibfnamefont {L.}~\bibnamefont
  {Li}}, \bibinfo {author} {\bibfnamefont {Y.}~\bibnamefont {Yu}}, \bibinfo
  {author} {\bibfnamefont {G.~J.}\ \bibnamefont {Ye}}, \bibinfo {author}
  {\bibfnamefont {Q.}~\bibnamefont {Ge}}, \bibinfo {author} {\bibfnamefont
  {X.}~\bibnamefont {Ou}}, \bibinfo {author} {\bibfnamefont {H.}~\bibnamefont
  {Wu}}, \bibinfo {author} {\bibfnamefont {D.}~\bibnamefont {Feng}}, \bibinfo
  {author} {\bibfnamefont {X.~H.}\ \bibnamefont {Chen}}, \ and\ \bibinfo
  {author} {\bibfnamefont {Y.}~\bibnamefont {Zhang}},\ }\href@noop {}
  {\bibfield  {journal} {\bibinfo  {journal} {Nat Nano}\ }\textbf {\bibinfo
  {volume} {9}},\ \bibinfo {pages} {372} (\bibinfo {year} {2014})}\BibitemShut
  {NoStop}%
\bibitem [{\citenamefont {Liu}\ \emph {et~al.}(2014{\natexlab{a}})\citenamefont
  {Liu}, \citenamefont {Neal}, \citenamefont {Zhu}, \citenamefont {Luo},
  \citenamefont {Xu}, \citenamefont {Tomanek},\ and\ \citenamefont
  {Ye}}]{10.1021/nn501226z}%
  \BibitemOpen
  \bibfield  {author} {\bibinfo {author} {\bibfnamefont {H.}~\bibnamefont
  {Liu}}, \bibinfo {author} {\bibfnamefont {A.~T.}\ \bibnamefont {Neal}},
  \bibinfo {author} {\bibfnamefont {Z.}~\bibnamefont {Zhu}}, \bibinfo {author}
  {\bibfnamefont {Z.}~\bibnamefont {Luo}}, \bibinfo {author} {\bibfnamefont
  {X.}~\bibnamefont {Xu}}, \bibinfo {author} {\bibfnamefont {D.}~\bibnamefont
  {Tomanek}}, \ and\ \bibinfo {author} {\bibfnamefont {P.~D.}\ \bibnamefont
  {Ye}},\ }\href@noop {} {\bibfield  {journal} {\bibinfo  {journal} {ACS Nano}\
  }\textbf {\bibinfo {volume} {8}},\ \bibinfo {pages} {4033} (\bibinfo {year}
  {2014}{\natexlab{a}})}\BibitemShut {NoStop}%
\bibitem [{\citenamefont {Pei}\ \emph {et~al.}(2016)\citenamefont {Pei},
  \citenamefont {Gai}, \citenamefont {Yang}, \citenamefont {Wang},
  \citenamefont {Yu}, \citenamefont {Choi}, \citenamefont {Luther-Davies},\
  and\ \citenamefont {Lu}}]{NatureCommun10450}%
  \BibitemOpen
  \bibfield  {author} {\bibinfo {author} {\bibfnamefont {J.}~\bibnamefont
  {Pei}}, \bibinfo {author} {\bibfnamefont {X.}~\bibnamefont {Gai}}, \bibinfo
  {author} {\bibfnamefont {J.}~\bibnamefont {Yang}}, \bibinfo {author}
  {\bibfnamefont {X.}~\bibnamefont {Wang}}, \bibinfo {author} {\bibfnamefont
  {Z.}~\bibnamefont {Yu}}, \bibinfo {author} {\bibfnamefont {D.-Y.}\
  \bibnamefont {Choi}}, \bibinfo {author} {\bibfnamefont {B.}~\bibnamefont
  {Luther-Davies}}, \ and\ \bibinfo {author} {\bibfnamefont {Y.}~\bibnamefont
  {Lu}},\ }\href@noop {} {\bibfield  {journal} {\bibinfo  {journal} {Nat.
  Commun.}\ }\textbf {\bibinfo {volume} {7}},\ \bibinfo {pages} {10450}
  (\bibinfo {year} {2016})}\BibitemShut {NoStop}%
\bibitem [{\citenamefont {Liang}\ \emph {et~al.}(2014)\citenamefont {Liang},
  \citenamefont {Wang}, \citenamefont {Lin}, \citenamefont {Sumpter},
  \citenamefont {Meunier},\ and\ \citenamefont {Pan}}]{10.1021/nl502892t}%
  \BibitemOpen
  \bibfield  {author} {\bibinfo {author} {\bibfnamefont {L.}~\bibnamefont
  {Liang}}, \bibinfo {author} {\bibfnamefont {J.}~\bibnamefont {Wang}},
  \bibinfo {author} {\bibfnamefont {W.}~\bibnamefont {Lin}}, \bibinfo {author}
  {\bibfnamefont {B.~G.}\ \bibnamefont {Sumpter}}, \bibinfo {author}
  {\bibfnamefont {V.}~\bibnamefont {Meunier}}, \ and\ \bibinfo {author}
  {\bibfnamefont {M.}~\bibnamefont {Pan}},\ }\href@noop {} {\bibfield
  {journal} {\bibinfo  {journal} {Nano Lett.}\ }\textbf {\bibinfo {volume}
  {14}},\ \bibinfo {pages} {6400} (\bibinfo {year} {2014})}\BibitemShut
  {NoStop}%
\bibitem [{\citenamefont {Qiao}\ \emph {et~al.}(2014)\citenamefont {Qiao},
  \citenamefont {Kong}, \citenamefont {Hu}, \citenamefont {Yang},\ and\
  \citenamefont {Ji}}]{10.1038/ncomms5475}%
  \BibitemOpen
  \bibfield  {author} {\bibinfo {author} {\bibfnamefont {J.}~\bibnamefont
  {Qiao}}, \bibinfo {author} {\bibfnamefont {X.}~\bibnamefont {Kong}}, \bibinfo
  {author} {\bibfnamefont {Z.-X.}\ \bibnamefont {Hu}}, \bibinfo {author}
  {\bibfnamefont {F.}~\bibnamefont {Yang}}, \ and\ \bibinfo {author}
  {\bibfnamefont {W.}~\bibnamefont {Ji}},\ }\href@noop {} {\bibfield  {journal}
  {\bibinfo  {journal} {Nat. Commun.}\ }\textbf {\bibinfo {volume} {5}},\
  \bibinfo {pages} {4475} (\bibinfo {year} {2014})}\BibitemShut {NoStop}%
\bibitem [{\citenamefont {Rodin}\ \emph {et~al.}(2014)\citenamefont {Rodin},
  \citenamefont {Carvalho},\ and\ \citenamefont
  {Castro~Neto}}]{PhysRevLett.112.176801}%
  \BibitemOpen
  \bibfield  {author} {\bibinfo {author} {\bibfnamefont {A.~S.}\ \bibnamefont
  {Rodin}}, \bibinfo {author} {\bibfnamefont {A.}~\bibnamefont {Carvalho}}, \
  and\ \bibinfo {author} {\bibfnamefont {A.~H.}\ \bibnamefont {Castro~Neto}},\
  }\href@noop {} {\bibfield  {journal} {\bibinfo  {journal} {Phys. Rev. Lett.}\
  }\textbf {\bibinfo {volume} {112}},\ \bibinfo {pages} {176801} (\bibinfo
  {year} {2014})}\BibitemShut {NoStop}%
\bibitem [{\citenamefont {Fei}\ \emph {et~al.}(2014)\citenamefont {Fei},
  \citenamefont {Faghaninia}, \citenamefont {Soklaski}, \citenamefont {Yan},
  \citenamefont {Lo},\ and\ \citenamefont {Yang}}]{10.1021/nl502865s}%
  \BibitemOpen
  \bibfield  {author} {\bibinfo {author} {\bibfnamefont {R.}~\bibnamefont
  {Fei}}, \bibinfo {author} {\bibfnamefont {A.}~\bibnamefont {Faghaninia}},
  \bibinfo {author} {\bibfnamefont {R.}~\bibnamefont {Soklaski}}, \bibinfo
  {author} {\bibfnamefont {J.-A.}\ \bibnamefont {Yan}}, \bibinfo {author}
  {\bibfnamefont {C.}~\bibnamefont {Lo}}, \ and\ \bibinfo {author}
  {\bibfnamefont {L.}~\bibnamefont {Yang}},\ }\href@noop {} {\bibfield
  {journal} {\bibinfo  {journal} {Nano Lett.}\ }\textbf {\bibinfo {volume}
  {14}},\ \bibinfo {pages} {6393} (\bibinfo {year} {2014})}\BibitemShut
  {NoStop}%
\bibitem [{\citenamefont {Sevincli}\ \emph {et~al.}(2008)\citenamefont
  {Sevincli}, \citenamefont {Topsakal}, \citenamefont {Durgun},\ and\
  \citenamefont {Ciraci}}]{PhysRevB.77.195434}%
  \BibitemOpen
  \bibfield  {author} {\bibinfo {author} {\bibfnamefont {H.}~\bibnamefont
  {Sevincli}}, \bibinfo {author} {\bibfnamefont {M.}~\bibnamefont {Topsakal}},
  \bibinfo {author} {\bibfnamefont {E.}~\bibnamefont {Durgun}}, \ and\ \bibinfo
  {author} {\bibfnamefont {S.}~\bibnamefont {Ciraci}},\ }\href@noop {}
  {\bibfield  {journal} {\bibinfo  {journal} {Phys. Rev. B}\ }\textbf {\bibinfo
  {volume} {77}},\ \bibinfo {pages} {195434} (\bibinfo {year}
  {2008})}\BibitemShut {NoStop}%
\bibitem [{\citenamefont {Krasheninnikov}\ \emph {et~al.}(2009)\citenamefont
  {Krasheninnikov}, \citenamefont {Lehtinen}, \citenamefont {Foster},
  \citenamefont {Pyykk\"o},\ and\ \citenamefont
  {Nieminen}}]{PhysRevLett.102.126807}%
  \BibitemOpen
  \bibfield  {author} {\bibinfo {author} {\bibfnamefont {A.~V.}\ \bibnamefont
  {Krasheninnikov}}, \bibinfo {author} {\bibfnamefont {P.~O.}\ \bibnamefont
  {Lehtinen}}, \bibinfo {author} {\bibfnamefont {A.~S.}\ \bibnamefont
  {Foster}}, \bibinfo {author} {\bibfnamefont {P.}~\bibnamefont {Pyykk\"o}}, \
  and\ \bibinfo {author} {\bibfnamefont {R.~M.}\ \bibnamefont {Nieminen}},\
  }\href@noop {} {\bibfield  {journal} {\bibinfo  {journal} {Phys. Rev. Lett.}\
  }\textbf {\bibinfo {volume} {102}},\ \bibinfo {pages} {126807} (\bibinfo
  {year} {2009})}\BibitemShut {NoStop}%
\bibitem [{\citenamefont {Han}\ \emph {et~al.}(2014)\citenamefont {Han},
  \citenamefont {Kawakami}, \citenamefont {Gmitra},\ and\ \citenamefont
  {Fabian}}]{10.1038/nnano.2014.214}%
  \BibitemOpen
  \bibfield  {author} {\bibinfo {author} {\bibfnamefont {W.}~\bibnamefont
  {Han}}, \bibinfo {author} {\bibfnamefont {R.~K.}\ \bibnamefont {Kawakami}},
  \bibinfo {author} {\bibfnamefont {M.}~\bibnamefont {Gmitra}}, \ and\ \bibinfo
  {author} {\bibfnamefont {J.}~\bibnamefont {Fabian}},\ }\href@noop {}
  {\bibfield  {journal} {\bibinfo  {journal} {Nat. Nano.}\ }\textbf {\bibinfo
  {volume} {9}},\ \bibinfo {pages} {794} (\bibinfo {year} {2014})}\BibitemShut
  {NoStop}%
\bibitem [{\citenamefont {Hu}\ and\ \citenamefont
  {Hong}(2015)}]{10.1021/acs.jpcc.5b01300}%
  \BibitemOpen
  \bibfield  {author} {\bibinfo {author} {\bibfnamefont {T.}~\bibnamefont
  {Hu}}\ and\ \bibinfo {author} {\bibfnamefont {J.}~\bibnamefont {Hong}},\
  }\href@noop {} {\bibfield  {journal} {\bibinfo  {journal} {J. Phys. Chem. C}\
  }\textbf {\bibinfo {volume} {119}},\ \bibinfo {pages} {8199} (\bibinfo {year}
  {2015})}\BibitemShut {NoStop}%
\bibitem [{\citenamefont {Hashmi}\ and\ \citenamefont
  {Hong}(2015)}]{10.1021/jp511574n}%
  \BibitemOpen
  \bibfield  {author} {\bibinfo {author} {\bibfnamefont {A.}~\bibnamefont
  {Hashmi}}\ and\ \bibinfo {author} {\bibfnamefont {J.}~\bibnamefont {Hong}},\
  }\href@noop {} {\bibfield  {journal} {\bibinfo  {journal} {J. Phys. Chem. C}\
  }\textbf {\bibinfo {volume} {119}},\ \bibinfo {pages} {9198} (\bibinfo {year}
  {2015})}\BibitemShut {NoStop}%
\bibitem [{\citenamefont {Sui}\ \emph {et~al.}(2015)\citenamefont {Sui},
  \citenamefont {Si}, \citenamefont {Shao}, \citenamefont {Zou}, \citenamefont
  {Wu}, \citenamefont {Gu},\ and\ \citenamefont {Duan}}]{10.1021/jp5129468}%
  \BibitemOpen
  \bibfield  {author} {\bibinfo {author} {\bibfnamefont {X.}~\bibnamefont
  {Sui}}, \bibinfo {author} {\bibfnamefont {C.}~\bibnamefont {Si}}, \bibinfo
  {author} {\bibfnamefont {B.}~\bibnamefont {Shao}}, \bibinfo {author}
  {\bibfnamefont {X.}~\bibnamefont {Zou}}, \bibinfo {author} {\bibfnamefont
  {J.}~\bibnamefont {Wu}}, \bibinfo {author} {\bibfnamefont {B.-L.}\
  \bibnamefont {Gu}}, \ and\ \bibinfo {author} {\bibfnamefont {W.}~\bibnamefont
  {Duan}},\ }\href@noop {} {\bibfield  {journal} {\bibinfo  {journal} {J. Phys.
  Chem. C}\ }\textbf {\bibinfo {volume} {119}},\ \bibinfo {pages} {10059}
  (\bibinfo {year} {2015})}\BibitemShut {NoStop}%
\bibitem [{\citenamefont {Seixas}\ \emph {et~al.}(2015)\citenamefont {Seixas},
  \citenamefont {Carvalho},\ and\ \citenamefont
  {Castro~Neto}}]{PhysRevB.91.155138}%
  \BibitemOpen
  \bibfield  {author} {\bibinfo {author} {\bibfnamefont {L.}~\bibnamefont
  {Seixas}}, \bibinfo {author} {\bibfnamefont {A.}~\bibnamefont {Carvalho}}, \
  and\ \bibinfo {author} {\bibfnamefont {A.~H.}\ \bibnamefont {Castro~Neto}},\
  }\href@noop {} {\bibfield  {journal} {\bibinfo  {journal} {Phys. Rev. B}\
  }\textbf {\bibinfo {volume} {91}},\ \bibinfo {pages} {155138} (\bibinfo
  {year} {2015})}\BibitemShut {NoStop}%
\bibitem [{\citenamefont {Lisenkov}\ \emph {et~al.}(2012)\citenamefont
  {Lisenkov}, \citenamefont {Andriotis},\ and\ \citenamefont
  {Menon}}]{PhysRevLett.108.187208}%
  \BibitemOpen
  \bibfield  {author} {\bibinfo {author} {\bibfnamefont {S.}~\bibnamefont
  {Lisenkov}}, \bibinfo {author} {\bibfnamefont {A.~N.}\ \bibnamefont
  {Andriotis}}, \ and\ \bibinfo {author} {\bibfnamefont {M.}~\bibnamefont
  {Menon}},\ }\href@noop {} {\bibfield  {journal} {\bibinfo  {journal} {Phys.
  Rev. Lett.}\ }\textbf {\bibinfo {volume} {108}},\ \bibinfo {pages} {187208}
  (\bibinfo {year} {2012})}\BibitemShut {NoStop}%
\bibitem [{\citenamefont {Gan}\ \emph {et~al.}(2008)\citenamefont {Gan},
  \citenamefont {Sun},\ and\ \citenamefont {Banhart}}]{SMLL200700929}%
  \BibitemOpen
  \bibfield  {author} {\bibinfo {author} {\bibfnamefont {Y.}~\bibnamefont
  {Gan}}, \bibinfo {author} {\bibfnamefont {L.}~\bibnamefont {Sun}}, \ and\
  \bibinfo {author} {\bibfnamefont {F.}~\bibnamefont {Banhart}},\ }\href@noop
  {} {\bibfield  {journal} {\bibinfo  {journal} {Small}\ }\textbf {\bibinfo
  {volume} {4}},\ \bibinfo {pages} {587} (\bibinfo {year} {2008})}\BibitemShut
  {NoStop}%
\bibitem [{\citenamefont {Krasheninnikov}\ and\ \citenamefont
  {Banhart}(2007)}]{10.1038/nmat1996}%
  \BibitemOpen
  \bibfield  {author} {\bibinfo {author} {\bibfnamefont {A.~V.}\ \bibnamefont
  {Krasheninnikov}}\ and\ \bibinfo {author} {\bibfnamefont {F.}~\bibnamefont
  {Banhart}},\ }\href@noop {} {\bibfield  {journal} {\bibinfo  {journal} {Nat.
  Mater.}\ }\textbf {\bibinfo {volume} {6}},\ \bibinfo {pages} {723} (\bibinfo
  {year} {2007})}\BibitemShut {NoStop}%
\bibitem [{\citenamefont {Wang}\ \emph
  {et~al.}(2012{\natexlab{b}})\citenamefont {Wang}, \citenamefont {Wang},
  \citenamefont {Cheng}, \citenamefont {Li}, \citenamefont {Yao}, \citenamefont
  {Zhang}, \citenamefont {Dong}, \citenamefont {Wang}, \citenamefont
  {Schwingenschlögl}, \citenamefont {Yang},\ and\ \citenamefont
  {Zhang}}]{10.1021/nl2031629}%
  \BibitemOpen
  \bibfield  {author} {\bibinfo {author} {\bibfnamefont {H.}~\bibnamefont
  {Wang}}, \bibinfo {author} {\bibfnamefont {Q.}~\bibnamefont {Wang}}, \bibinfo
  {author} {\bibfnamefont {Y.}~\bibnamefont {Cheng}}, \bibinfo {author}
  {\bibfnamefont {K.}~\bibnamefont {Li}}, \bibinfo {author} {\bibfnamefont
  {Y.}~\bibnamefont {Yao}}, \bibinfo {author} {\bibfnamefont {Q.}~\bibnamefont
  {Zhang}}, \bibinfo {author} {\bibfnamefont {C.}~\bibnamefont {Dong}},
  \bibinfo {author} {\bibfnamefont {P.}~\bibnamefont {Wang}}, \bibinfo {author}
  {\bibfnamefont {U.}~\bibnamefont {Schwingenschlögl}}, \bibinfo {author}
  {\bibfnamefont {W.}~\bibnamefont {Yang}}, \ and\ \bibinfo {author}
  {\bibfnamefont {X.~X.}\ \bibnamefont {Zhang}},\ }\href@noop {} {\bibfield
  {journal} {\bibinfo  {journal} {Nano Lett.}\ }\textbf {\bibinfo {volume}
  {12}},\ \bibinfo {pages} {141} (\bibinfo {year}
  {2012}{\natexlab{b}})}\BibitemShut {NoStop}%
\bibitem [{\citenamefont {Kresse}\ and\ \citenamefont
  {Hafner}(1993)}]{PhysRevB.47.558}%
  \BibitemOpen
  \bibfield  {author} {\bibinfo {author} {\bibfnamefont {G.}~\bibnamefont
  {Kresse}}\ and\ \bibinfo {author} {\bibfnamefont {J.}~\bibnamefont
  {Hafner}},\ }\href@noop {} {\bibfield  {journal} {\bibinfo  {journal} {Phys.
  Rev. B}\ }\textbf {\bibinfo {volume} {47}},\ \bibinfo {pages} {558} (\bibinfo
  {year} {1993})}\BibitemShut {NoStop}%
\bibitem [{\citenamefont {Kresse}\ and\ \citenamefont
  {Furthm\"uller}(1996)}]{PhysRevB.54.11169}%
  \BibitemOpen
  \bibfield  {author} {\bibinfo {author} {\bibfnamefont {G.}~\bibnamefont
  {Kresse}}\ and\ \bibinfo {author} {\bibfnamefont {J.}~\bibnamefont
  {Furthm\"uller}},\ }\href@noop {} {\bibfield  {journal} {\bibinfo  {journal}
  {Phys. Rev. B}\ }\textbf {\bibinfo {volume} {54}},\ \bibinfo {pages} {11169}
  (\bibinfo {year} {1996})}\BibitemShut {NoStop}%
\bibitem [{\citenamefont {Bl\"ochl}(1994)}]{PhysRevB.50.17953}%
  \BibitemOpen
  \bibfield  {author} {\bibinfo {author} {\bibfnamefont {P.~E.}\ \bibnamefont
  {Bl\"ochl}},\ }\href@noop {} {\bibfield  {journal} {\bibinfo  {journal}
  {Phys. Rev. B}\ }\textbf {\bibinfo {volume} {50}},\ \bibinfo {pages} {17953}
  (\bibinfo {year} {1994})}\BibitemShut {NoStop}%
\bibitem [{\citenamefont {Perdew}\ \emph {et~al.}(1996)\citenamefont {Perdew},
  \citenamefont {Burke},\ and\ \citenamefont
  {Ernzerhof}}]{PhysRevLett.77.3865}%
  \BibitemOpen
  \bibfield  {author} {\bibinfo {author} {\bibfnamefont {J.~P.}\ \bibnamefont
  {Perdew}}, \bibinfo {author} {\bibfnamefont {K.}~\bibnamefont {Burke}}, \
  and\ \bibinfo {author} {\bibfnamefont {M.}~\bibnamefont {Ernzerhof}},\
  }\href@noop {} {\bibfield  {journal} {\bibinfo  {journal} {Phys. Rev. Lett.}\
  }\textbf {\bibinfo {volume} {77}},\ \bibinfo {pages} {3865} (\bibinfo {year}
  {1996})}\BibitemShut {NoStop}%
\bibitem [{\citenamefont {Heyd}\ \emph {et~al.}(2003)\citenamefont {Heyd},
  \citenamefont {Scuseria},\ and\ \citenamefont {Ernzerhof}}]{jcp1.1564060}%
  \BibitemOpen
  \bibfield  {author} {\bibinfo {author} {\bibfnamefont {J.}~\bibnamefont
  {Heyd}}, \bibinfo {author} {\bibfnamefont {G.~E.}\ \bibnamefont {Scuseria}},
  \ and\ \bibinfo {author} {\bibfnamefont {M.}~\bibnamefont {Ernzerhof}},\
  }\href@noop {} {\bibfield  {journal} {\bibinfo  {journal} {J. Chem. Phys.}\
  }\textbf {\bibinfo {volume} {118}},\ \bibinfo {pages} {8207} (\bibinfo {year}
  {2003})}\BibitemShut {NoStop}%
\bibitem [{\citenamefont {Henkelman}\ \emph {et~al.}(2000)\citenamefont
  {Henkelman}, \citenamefont {Uberuaga},\ and\ \citenamefont
  {Jónsson}}]{1.1329672}%
  \BibitemOpen
  \bibfield  {author} {\bibinfo {author} {\bibfnamefont {G.}~\bibnamefont
  {Henkelman}}, \bibinfo {author} {\bibfnamefont {B.~P.}\ \bibnamefont
  {Uberuaga}}, \ and\ \bibinfo {author} {\bibfnamefont {H.}~\bibnamefont
  {Jónsson}},\ }\href@noop {} {\bibfield  {journal} {\bibinfo  {journal} {J.
  Chem. Phys.}\ }\textbf {\bibinfo {volume} {113}},\ \bibinfo {pages} {9901}
  (\bibinfo {year} {2000})}\BibitemShut {NoStop}%
\bibitem [{\citenamefont {Dion}\ \emph {et~al.}(2004)\citenamefont {Dion},
  \citenamefont {Rydberg}, \citenamefont {Schr\"oder}, \citenamefont
  {Langreth},\ and\ \citenamefont {Lundqvist}}]{PhysRevLett.92.246401}%
  \BibitemOpen
  \bibfield  {author} {\bibinfo {author} {\bibfnamefont {M.}~\bibnamefont
  {Dion}}, \bibinfo {author} {\bibfnamefont {H.}~\bibnamefont {Rydberg}},
  \bibinfo {author} {\bibfnamefont {E.}~\bibnamefont {Schr\"oder}}, \bibinfo
  {author} {\bibfnamefont {D.~C.}\ \bibnamefont {Langreth}}, \ and\ \bibinfo
  {author} {\bibfnamefont {B.~I.}\ \bibnamefont {Lundqvist}},\ }\href@noop {}
  {\bibfield  {journal} {\bibinfo  {journal} {Phys. Rev. Lett.}\ }\textbf
  {\bibinfo {volume} {92}},\ \bibinfo {pages} {246401} (\bibinfo {year}
  {2004})}\BibitemShut {NoStop}%
\bibitem [{\citenamefont {Klimes}\ \emph {et~al.}(2011)\citenamefont {Klimes},
  \citenamefont {Bowler},\ and\ \citenamefont
  {Michaelides}}]{PhysRevB.83.195131}%
  \BibitemOpen
  \bibfield  {author} {\bibinfo {author} {\bibfnamefont {J.}~\bibnamefont
  {Klimes}}, \bibinfo {author} {\bibfnamefont {D.~R.}\ \bibnamefont {Bowler}},
  \ and\ \bibinfo {author} {\bibfnamefont {A.}~\bibnamefont {Michaelides}},\
  }\href@noop {} {\bibfield  {journal} {\bibinfo  {journal} {Phys. Rev. B}\
  }\textbf {\bibinfo {volume} {83}},\ \bibinfo {pages} {195131} (\bibinfo
  {year} {2011})}\BibitemShut {NoStop}%
\bibitem [{\citenamefont {Brown}\ and\ \citenamefont
  {Rundqvist}(1965)}]{10.1107/S0365110X65004140}%
  \BibitemOpen
  \bibfield  {author} {\bibinfo {author} {\bibfnamefont {A.}~\bibnamefont
  {Brown}}\ and\ \bibinfo {author} {\bibfnamefont {S.}~\bibnamefont
  {Rundqvist}},\ }\href@noop {} {\bibfield  {journal} {\bibinfo  {journal}
  {Acta Cryst.}\ }\textbf {\bibinfo {volume} {19}},\ \bibinfo {pages} {684}
  (\bibinfo {year} {1965})}\BibitemShut {NoStop}%
\bibitem [{\citenamefont {Becke}\ and\ \citenamefont
  {Johnson}(2006)}]{10.1063/1.2213970}%
  \BibitemOpen
  \bibfield  {author} {\bibinfo {author} {\bibfnamefont {A.~D.}\ \bibnamefont
  {Becke}}\ and\ \bibinfo {author} {\bibfnamefont {E.~R.}\ \bibnamefont
  {Johnson}},\ }\href@noop {} {\bibfield  {journal} {\bibinfo  {journal} {J.
  Chem. Phys.}\ }\textbf {\bibinfo {volume} {124}},\ \bibinfo {pages} {221101}
  (\bibinfo {year} {2006})}\BibitemShut {NoStop}%
\bibitem [{\citenamefont {Tran}\ and\ \citenamefont
  {Blaha}(2009)}]{PhysRevLett.102.226401}%
  \BibitemOpen
  \bibfield  {author} {\bibinfo {author} {\bibfnamefont {F.}~\bibnamefont
  {Tran}}\ and\ \bibinfo {author} {\bibfnamefont {P.}~\bibnamefont {Blaha}},\
  }\href@noop {} {\bibfield  {journal} {\bibinfo  {journal} {Phys. Rev. Lett.}\
  }\textbf {\bibinfo {volume} {102}},\ \bibinfo {pages} {226401} (\bibinfo
  {year} {2009})}\BibitemShut {NoStop}%
\bibitem [{\citenamefont {Keyes}(1953)}]{PhysRev.92.580}%
  \BibitemOpen
  \bibfield  {author} {\bibinfo {author} {\bibfnamefont {R.~W.}\ \bibnamefont
  {Keyes}},\ }\href@noop {} {\bibfield  {journal} {\bibinfo  {journal} {Phys.
  Rev.}\ }\textbf {\bibinfo {volume} {92}},\ \bibinfo {pages} {580} (\bibinfo
  {year} {1953})}\BibitemShut {NoStop}%
\bibitem [{\citenamefont {Warschauer}(1963)}]{10.1063/1.1729699}%
  \BibitemOpen
  \bibfield  {author} {\bibinfo {author} {\bibfnamefont {D.}~\bibnamefont
  {Warschauer}},\ }\href@noop {} {\bibfield  {journal} {\bibinfo  {journal} {J.
  App. Phys.}\ }\textbf {\bibinfo {volume} {34}},\ \bibinfo {pages} {1853}
  (\bibinfo {year} {1963})}\BibitemShut {NoStop}%
\bibitem [{\citenamefont {Maruyama}\ \emph {et~al.}(1981)\citenamefont
  {Maruyama}, \citenamefont {Suzuki}, \citenamefont {Kobayashi},\ and\
  \citenamefont {Tanuma}}]{Maruyama198199}%
  \BibitemOpen
  \bibfield  {author} {\bibinfo {author} {\bibfnamefont {Y.}~\bibnamefont
  {Maruyama}}, \bibinfo {author} {\bibfnamefont {S.}~\bibnamefont {Suzuki}},
  \bibinfo {author} {\bibfnamefont {K.}~\bibnamefont {Kobayashi}}, \ and\
  \bibinfo {author} {\bibfnamefont {S.}~\bibnamefont {Tanuma}},\ }\href@noop {}
  {\bibfield  {journal} {\bibinfo  {journal} {Physica B+C}\ }\textbf {\bibinfo
  {volume} {105}},\ \bibinfo {pages} {99 } (\bibinfo {year}
  {1981})}\BibitemShut {NoStop}%
\bibitem [{\citenamefont {Rudenko}\ and\ \citenamefont
  {Katsnelson}(2014)}]{PhysRevB.89.201408}%
  \BibitemOpen
  \bibfield  {author} {\bibinfo {author} {\bibfnamefont {A.~N.}\ \bibnamefont
  {Rudenko}}\ and\ \bibinfo {author} {\bibfnamefont {M.~I.}\ \bibnamefont
  {Katsnelson}},\ }\href {\doibase 10.1103/PhysRevB.89.201408} {\bibfield
  {journal} {\bibinfo  {journal} {Phys. Rev. B}\ }\textbf {\bibinfo {volume}
  {89}},\ \bibinfo {pages} {201408} (\bibinfo {year} {2014})}\BibitemShut
  {NoStop}%
\bibitem [{\citenamefont {Srivastava}\ \emph {et~al.}(2015)\citenamefont
  {Srivastava}, \citenamefont {Hembram}, \citenamefont {Mizuseki},
  \citenamefont {Lee}, \citenamefont {Han},\ and\ \citenamefont
  {Kim}}]{10.1021/jp5110938}%
  \BibitemOpen
  \bibfield  {author} {\bibinfo {author} {\bibfnamefont {P.}~\bibnamefont
  {Srivastava}}, \bibinfo {author} {\bibfnamefont {K.~P. S.~S.}\ \bibnamefont
  {Hembram}}, \bibinfo {author} {\bibfnamefont {H.}~\bibnamefont {Mizuseki}},
  \bibinfo {author} {\bibfnamefont {K.-R.}\ \bibnamefont {Lee}}, \bibinfo
  {author} {\bibfnamefont {S.~S.}\ \bibnamefont {Han}}, \ and\ \bibinfo
  {author} {\bibfnamefont {S.}~\bibnamefont {Kim}},\ }\href@noop {} {\bibfield
  {journal} {\bibinfo  {journal} {J. Phys. Chem. C}\ }\textbf {\bibinfo
  {volume} {119}},\ \bibinfo {pages} {6530} (\bibinfo {year}
  {2015})}\BibitemShut {NoStop}%
\bibitem [{\citenamefont {Liu}\ \emph {et~al.}(2014{\natexlab{b}})\citenamefont
  {Liu}, \citenamefont {Xu}, \citenamefont {Zhang}, \citenamefont {Penev},\
  and\ \citenamefont {Yakobson}}]{10.1021/nl5021393}%
  \BibitemOpen
  \bibfield  {author} {\bibinfo {author} {\bibfnamefont {Y.}~\bibnamefont
  {Liu}}, \bibinfo {author} {\bibfnamefont {F.}~\bibnamefont {Xu}}, \bibinfo
  {author} {\bibfnamefont {Z.}~\bibnamefont {Zhang}}, \bibinfo {author}
  {\bibfnamefont {E.~S.}\ \bibnamefont {Penev}}, \ and\ \bibinfo {author}
  {\bibfnamefont {B.~I.}\ \bibnamefont {Yakobson}},\ }\href@noop {} {\bibfield
  {journal} {\bibinfo  {journal} {Nano Lett.}\ }\textbf {\bibinfo {volume}
  {14}},\ \bibinfo {pages} {6782} (\bibinfo {year}
  {2014}{\natexlab{b}})}\BibitemShut {NoStop}%
\bibitem [{\citenamefont {Banhart}\ \emph {et~al.}(2011)\citenamefont
  {Banhart}, \citenamefont {Kotakoski},\ and\ \citenamefont
  {Krasheninnikov}}]{10.1021/nn102598m}%
  \BibitemOpen
  \bibfield  {author} {\bibinfo {author} {\bibfnamefont {F.}~\bibnamefont
  {Banhart}}, \bibinfo {author} {\bibfnamefont {J.}~\bibnamefont {Kotakoski}},
  \ and\ \bibinfo {author} {\bibfnamefont {A.~V.}\ \bibnamefont
  {Krasheninnikov}},\ }\href@noop {} {\bibfield  {journal} {\bibinfo  {journal}
  {ACS Nano}\ }\textbf {\bibinfo {volume} {5}},\ \bibinfo {pages} {26}
  (\bibinfo {year} {2011})}\BibitemShut {NoStop}%
\bibitem [{\citenamefont {Komsa}\ \emph {et~al.}(2012)\citenamefont {Komsa},
  \citenamefont {Kotakoski}, \citenamefont {Kurasch}, \citenamefont {Lehtinen},
  \citenamefont {Kaiser},\ and\ \citenamefont
  {Krasheninnikov}}]{PhysRevLett.109.035503}%
  \BibitemOpen
  \bibfield  {author} {\bibinfo {author} {\bibfnamefont {H.-P.}\ \bibnamefont
  {Komsa}}, \bibinfo {author} {\bibfnamefont {J.}~\bibnamefont {Kotakoski}},
  \bibinfo {author} {\bibfnamefont {S.}~\bibnamefont {Kurasch}}, \bibinfo
  {author} {\bibfnamefont {O.}~\bibnamefont {Lehtinen}}, \bibinfo {author}
  {\bibfnamefont {U.}~\bibnamefont {Kaiser}}, \ and\ \bibinfo {author}
  {\bibfnamefont {A.~V.}\ \bibnamefont {Krasheninnikov}},\ }\href@noop {}
  {\bibfield  {journal} {\bibinfo  {journal} {Phys. Rev. Lett.}\ }\textbf
  {\bibinfo {volume} {109}},\ \bibinfo {pages} {035503} (\bibinfo {year}
  {2012})}\BibitemShut {NoStop}%
\bibitem [{\citenamefont {Palacios}\ \emph {et~al.}(2008)\citenamefont
  {Palacios}, \citenamefont {Fern\'andez-Rossier},\ and\ \citenamefont
  {Brey}}]{PhysRevB.77.195428}%
  \BibitemOpen
  \bibfield  {author} {\bibinfo {author} {\bibfnamefont {J.~J.}\ \bibnamefont
  {Palacios}}, \bibinfo {author} {\bibfnamefont {J.}~\bibnamefont
  {Fern\'andez-Rossier}}, \ and\ \bibinfo {author} {\bibfnamefont
  {L.}~\bibnamefont {Brey}},\ }\href@noop {} {\bibfield  {journal} {\bibinfo
  {journal} {Phys. Rev. B}\ }\textbf {\bibinfo {volume} {77}},\ \bibinfo
  {pages} {195428} (\bibinfo {year} {2008})}\BibitemShut {NoStop}%
\bibitem [{\citenamefont {Attaccalite}\ \emph {et~al.}(2011)\citenamefont
  {Attaccalite}, \citenamefont {Bockstedte}, \citenamefont {Marini},
  \citenamefont {Rubio},\ and\ \citenamefont {Wirtz}}]{PhysRevB.83.144115}%
  \BibitemOpen
  \bibfield  {author} {\bibinfo {author} {\bibfnamefont {C.}~\bibnamefont
  {Attaccalite}}, \bibinfo {author} {\bibfnamefont {M.}~\bibnamefont
  {Bockstedte}}, \bibinfo {author} {\bibfnamefont {A.}~\bibnamefont {Marini}},
  \bibinfo {author} {\bibfnamefont {A.}~\bibnamefont {Rubio}}, \ and\ \bibinfo
  {author} {\bibfnamefont {L.}~\bibnamefont {Wirtz}},\ }\href@noop {}
  {\bibfield  {journal} {\bibinfo  {journal} {Phys. Rev. B}\ }\textbf {\bibinfo
  {volume} {83}},\ \bibinfo {pages} {144115} (\bibinfo {year}
  {2011})}\BibitemShut {NoStop}%
\bibitem [{\citenamefont {Hu}\ and\ \citenamefont
  {Yang}(2015)}]{10.1021/acs.jpcc.5b06077}%
  \BibitemOpen
  \bibfield  {author} {\bibinfo {author} {\bibfnamefont {W.}~\bibnamefont
  {Hu}}\ and\ \bibinfo {author} {\bibfnamefont {J.}~\bibnamefont {Yang}},\
  }\href@noop {} {\bibfield  {journal} {\bibinfo  {journal} {J. Phys. Chem. C}\
  }\textbf {\bibinfo {volume} {119}},\ \bibinfo {pages} {20474} (\bibinfo
  {year} {2015})}\BibitemShut {NoStop}%
\bibitem [{\citenamefont {Hu}\ and\ \citenamefont
  {Dong}(2015)}]{0957-4484-26-6-065705}%
  \BibitemOpen
  \bibfield  {author} {\bibinfo {author} {\bibfnamefont {T.}~\bibnamefont
  {Hu}}\ and\ \bibinfo {author} {\bibfnamefont {J.}~\bibnamefont {Dong}},\
  }\href@noop {} {\bibfield  {journal} {\bibinfo  {journal} {Nanotechnology}\
  }\textbf {\bibinfo {volume} {26}},\ \bibinfo {pages} {065705} (\bibinfo
  {year} {2015})}\BibitemShut {NoStop}%
\bibitem [{\citenamefont {El-Barbary}\ \emph {et~al.}(2003)\citenamefont
  {El-Barbary}, \citenamefont {Telling}, \citenamefont {Ewels}, \citenamefont
  {Heggie},\ and\ \citenamefont {Briddon}}]{PhysRevB.68.144107}%
  \BibitemOpen
  \bibfield  {author} {\bibinfo {author} {\bibfnamefont {A.~A.}\ \bibnamefont
  {El-Barbary}}, \bibinfo {author} {\bibfnamefont {R.~H.}\ \bibnamefont
  {Telling}}, \bibinfo {author} {\bibfnamefont {C.~P.}\ \bibnamefont {Ewels}},
  \bibinfo {author} {\bibfnamefont {M.~I.}\ \bibnamefont {Heggie}}, \ and\
  \bibinfo {author} {\bibfnamefont {P.~R.}\ \bibnamefont {Briddon}},\
  }\href@noop {} {\bibfield  {journal} {\bibinfo  {journal} {Phys. Rev. B}\
  }\textbf {\bibinfo {volume} {68}},\ \bibinfo {pages} {144107} (\bibinfo
  {year} {2003})}\BibitemShut {NoStop}%
\bibitem [{\citenamefont {Hashimoto}\ \emph {et~al.}(2004)\citenamefont
  {Hashimoto}, \citenamefont {Suenaga}, \citenamefont {Gloter}, \citenamefont
  {Urita},\ and\ \citenamefont {Iijima}}]{10.1038/nature02817}%
  \BibitemOpen
  \bibfield  {author} {\bibinfo {author} {\bibfnamefont {A.}~\bibnamefont
  {Hashimoto}}, \bibinfo {author} {\bibfnamefont {K.}~\bibnamefont {Suenaga}},
  \bibinfo {author} {\bibfnamefont {A.}~\bibnamefont {Gloter}}, \bibinfo
  {author} {\bibfnamefont {K.}~\bibnamefont {Urita}}, \ and\ \bibinfo {author}
  {\bibfnamefont {S.}~\bibnamefont {Iijima}},\ }\href@noop {} {\bibfield
  {journal} {\bibinfo  {journal} {Nature}\ }\textbf {\bibinfo {volume} {430}},\
  \bibinfo {pages} {870} (\bibinfo {year} {2004})}\BibitemShut {NoStop}%
\bibitem [{\citenamefont {Lee}\ \emph {et~al.}(2005)\citenamefont {Lee},
  \citenamefont {Wang}, \citenamefont {Yoon}, \citenamefont {Hwang},
  \citenamefont {Kim},\ and\ \citenamefont {Ho}}]{PhysRevLett.95.205501}%
  \BibitemOpen
  \bibfield  {author} {\bibinfo {author} {\bibfnamefont {G.-D.}\ \bibnamefont
  {Lee}}, \bibinfo {author} {\bibfnamefont {C.~Z.}\ \bibnamefont {Wang}},
  \bibinfo {author} {\bibfnamefont {E.}~\bibnamefont {Yoon}}, \bibinfo {author}
  {\bibfnamefont {N.-M.}\ \bibnamefont {Hwang}}, \bibinfo {author}
  {\bibfnamefont {D.-Y.}\ \bibnamefont {Kim}}, \ and\ \bibinfo {author}
  {\bibfnamefont {K.~M.}\ \bibnamefont {Ho}},\ }\href@noop {} {\bibfield
  {journal} {\bibinfo  {journal} {Phys. Rev. Lett.}\ }\textbf {\bibinfo
  {volume} {95}},\ \bibinfo {pages} {205501} (\bibinfo {year}
  {2005})}\BibitemShut {NoStop}%
\bibitem [{\citenamefont {Robertson}\ \emph {et~al.}(2013)\citenamefont
  {Robertson}, \citenamefont {Montanari}, \citenamefont {He}, \citenamefont
  {Allen}, \citenamefont {Wu}, \citenamefont {Harrison}, \citenamefont
  {Kirkland},\ and\ \citenamefont {Warner}}]{10.1021/nn401113r}%
  \BibitemOpen
  \bibfield  {author} {\bibinfo {author} {\bibfnamefont {A.~W.}\ \bibnamefont
  {Robertson}}, \bibinfo {author} {\bibfnamefont {B.}~\bibnamefont
  {Montanari}}, \bibinfo {author} {\bibfnamefont {K.}~\bibnamefont {He}},
  \bibinfo {author} {\bibfnamefont {C.~S.}\ \bibnamefont {Allen}}, \bibinfo
  {author} {\bibfnamefont {Y.~A.}\ \bibnamefont {Wu}}, \bibinfo {author}
  {\bibfnamefont {N.~M.}\ \bibnamefont {Harrison}}, \bibinfo {author}
  {\bibfnamefont {A.~I.}\ \bibnamefont {Kirkland}}, \ and\ \bibinfo {author}
  {\bibfnamefont {J.~H.}\ \bibnamefont {Warner}},\ }\href@noop {} {\bibfield
  {journal} {\bibinfo  {journal} {ACS Nano}\ }\textbf {\bibinfo {volume} {7}},\
  \bibinfo {pages} {4495} (\bibinfo {year} {2013})}\BibitemShut {NoStop}%
\bibitem [{\citenamefont {Gao}\ \emph {et~al.}(2013)\citenamefont {Gao},
  \citenamefont {Zhang}, \citenamefont {Liu}, \citenamefont {Zhang},\ and\
  \citenamefont {Zhao}}]{C3NR02826G}%
  \BibitemOpen
  \bibfield  {author} {\bibinfo {author} {\bibfnamefont {J.}~\bibnamefont
  {Gao}}, \bibinfo {author} {\bibfnamefont {J.}~\bibnamefont {Zhang}}, \bibinfo
  {author} {\bibfnamefont {H.}~\bibnamefont {Liu}}, \bibinfo {author}
  {\bibfnamefont {Q.}~\bibnamefont {Zhang}}, \ and\ \bibinfo {author}
  {\bibfnamefont {J.}~\bibnamefont {Zhao}},\ }\href@noop {} {\bibfield
  {journal} {\bibinfo  {journal} {Nanoscale}\ }\textbf {\bibinfo {volume}
  {5}},\ \bibinfo {pages} {9785} (\bibinfo {year} {2013})}\BibitemShut
  {NoStop}%
\bibitem [{\citenamefont {Deringer}\ \emph {et~al.}(2011)\citenamefont
  {Deringer}, \citenamefont {Tchougreeff},\ and\ \citenamefont
  {Dronskowski}}]{doi:10.1021/jp202489s}%
  \BibitemOpen
  \bibfield  {author} {\bibinfo {author} {\bibfnamefont {V.~L.}\ \bibnamefont
  {Deringer}}, \bibinfo {author} {\bibfnamefont {A.~L.}\ \bibnamefont
  {Tchougreeff}}, \ and\ \bibinfo {author} {\bibfnamefont {R.}~\bibnamefont
  {Dronskowski}},\ }\href@noop {} {\bibfield  {journal} {\bibinfo  {journal}
  {J. Phys. Chem. A}\ }\textbf {\bibinfo {volume} {115}},\ \bibinfo {pages}
  {5461} (\bibinfo {year} {2011})}\BibitemShut {NoStop}%
\bibitem [{\citenamefont {Dronskowski}\ and\ \citenamefont
  {Bloechl}(1993)}]{doi:10.1021/j100135a014}%
  \BibitemOpen
  \bibfield  {author} {\bibinfo {author} {\bibfnamefont {R.}~\bibnamefont
  {Dronskowski}}\ and\ \bibinfo {author} {\bibfnamefont {P.~E.}\ \bibnamefont
  {Bloechl}},\ }\href@noop {} {\bibfield  {journal} {\bibinfo  {journal} {J.
  Phys. Chem.}\ }\textbf {\bibinfo {volume} {97}},\ \bibinfo {pages} {8617}
  (\bibinfo {year} {1993})}\BibitemShut {NoStop}%
\bibitem [{\citenamefont {Pyykkö}\ and\ \citenamefont
  {Atsumi}(2009)}]{CHEM:CHEM200800987}%
  \BibitemOpen
  \bibfield  {author} {\bibinfo {author} {\bibfnamefont {P.}~\bibnamefont
  {Pyykkö}}\ and\ \bibinfo {author} {\bibfnamefont {M.}~\bibnamefont
  {Atsumi}},\ }\href@noop {} {\bibfield  {journal} {\bibinfo  {journal} {Chem.
  Eur. J.}\ }\textbf {\bibinfo {volume} {15}},\ \bibinfo {pages} {186}
  (\bibinfo {year} {2009})}\BibitemShut {NoStop}%
\bibitem [{\citenamefont {Koenig}\ \emph {et~al.}(2016)\citenamefont {Koenig},
  \citenamefont {Doganov}, \citenamefont {Seixas}, \citenamefont {Carvalho},
  \citenamefont {Tan}, \citenamefont {Watanabe}, \citenamefont {Taniguchi},
  \citenamefont {Yakovlev}, \citenamefont {Neto},\ and\ \citenamefont
  {Özyilmaz}}]{acs.nanolett.5b03278}%
  \BibitemOpen
  \bibfield  {author} {\bibinfo {author} {\bibfnamefont {S.~P.}\ \bibnamefont
  {Koenig}}, \bibinfo {author} {\bibfnamefont {R.~A.}\ \bibnamefont {Doganov}},
  \bibinfo {author} {\bibfnamefont {L.}~\bibnamefont {Seixas}}, \bibinfo
  {author} {\bibfnamefont {A.}~\bibnamefont {Carvalho}}, \bibinfo {author}
  {\bibfnamefont {J.~Y.}\ \bibnamefont {Tan}}, \bibinfo {author} {\bibfnamefont
  {K.}~\bibnamefont {Watanabe}}, \bibinfo {author} {\bibfnamefont
  {T.}~\bibnamefont {Taniguchi}}, \bibinfo {author} {\bibfnamefont
  {N.}~\bibnamefont {Yakovlev}}, \bibinfo {author} {\bibfnamefont {A.~H.~C.}\
  \bibnamefont {Neto}}, \ and\ \bibinfo {author} {\bibfnamefont
  {B.}~\bibnamefont {Özyilmaz}},\ }\href@noop {} {\bibfield  {journal}
  {\bibinfo  {journal} {Nano Lett.}\ }\textbf {\bibinfo {volume} {16}},\
  \bibinfo {pages} {2145} (\bibinfo {year} {2016})}\BibitemShut {NoStop}%
\bibitem [{\citenamefont {Sun}\ \emph {et~al.}(2015)\citenamefont {Sun},
  \citenamefont {Wang}, \citenamefont {Lin}, \citenamefont {Hou},\ and\
  \citenamefont {Li}}]{1.4921699}%
  \BibitemOpen
  \bibfield  {author} {\bibinfo {author} {\bibfnamefont {X.}~\bibnamefont
  {Sun}}, \bibinfo {author} {\bibfnamefont {L.}~\bibnamefont {Wang}}, \bibinfo
  {author} {\bibfnamefont {H.}~\bibnamefont {Lin}}, \bibinfo {author}
  {\bibfnamefont {T.}~\bibnamefont {Hou}}, \ and\ \bibinfo {author}
  {\bibfnamefont {Y.}~\bibnamefont {Li}},\ }\href@noop {} {\bibfield  {journal}
  {\bibinfo  {journal} {Appl. Phys. Lett.}\ }\textbf {\bibinfo {volume}
  {106}},\ \bibinfo {eid} {222401} (\bibinfo {year} {2015})}\BibitemShut
  {NoStop}%
\bibitem [{\citenamefont {Himpsel}\ \emph {et~al.}(1998)\citenamefont
  {Himpsel}, \citenamefont {Ortega}, \citenamefont {Mankey},\ and\
  \citenamefont {Willis}}]{doi:10.1080}%
  \BibitemOpen
  \bibfield  {author} {\bibinfo {author} {\bibfnamefont {F.~J.}\ \bibnamefont
  {Himpsel}}, \bibinfo {author} {\bibfnamefont {J.~E.}\ \bibnamefont {Ortega}},
  \bibinfo {author} {\bibfnamefont {G.~J.}\ \bibnamefont {Mankey}}, \ and\
  \bibinfo {author} {\bibfnamefont {R.~F.}\ \bibnamefont {Willis}},\
  }\href@noop {} {\bibfield  {journal} {\bibinfo  {journal} {Advances in
  Physics}\ }\textbf {\bibinfo {volume} {47}},\ \bibinfo {pages} {511}
  (\bibinfo {year} {1998})}\BibitemShut {NoStop}%
\bibitem [{\citenamefont {Himpsel}(1991)}]{PhysRevLett.67.2363}%
  \BibitemOpen
  \bibfield  {author} {\bibinfo {author} {\bibfnamefont {F.~J.}\ \bibnamefont
  {Himpsel}},\ }\href@noop {} {\bibfield  {journal} {\bibinfo  {journal} {Phys.
  Rev. Lett.}\ }\textbf {\bibinfo {volume} {67}},\ \bibinfo {pages} {2363}
  (\bibinfo {year} {1991})}\BibitemShut {NoStop}%
\bibitem [{\citenamefont {Ortega}\ and\ \citenamefont
  {Himpsel}(1993)}]{PhysRevB.47.16441}%
  \BibitemOpen
  \bibfield  {author} {\bibinfo {author} {\bibfnamefont {J.~E.}\ \bibnamefont
  {Ortega}}\ and\ \bibinfo {author} {\bibfnamefont {F.~J.}\ \bibnamefont
  {Himpsel}},\ }\href@noop {} {\bibfield  {journal} {\bibinfo  {journal} {Phys.
  Rev. B}\ }\textbf {\bibinfo {volume} {47}},\ \bibinfo {pages} {16441}
  (\bibinfo {year} {1993})}\BibitemShut {NoStop}%
\bibitem [{\citenamefont {Alkemper}\ \emph {et~al.}(1994)\citenamefont
  {Alkemper}, \citenamefont {Carbone}, \citenamefont {Vescovo}, \citenamefont
  {Eberhardt}, \citenamefont {Rader},\ and\ \citenamefont
  {Gudat}}]{PhysRevB.50.17496}%
  \BibitemOpen
  \bibfield  {author} {\bibinfo {author} {\bibfnamefont {U.}~\bibnamefont
  {Alkemper}}, \bibinfo {author} {\bibfnamefont {C.}~\bibnamefont {Carbone}},
  \bibinfo {author} {\bibfnamefont {E.}~\bibnamefont {Vescovo}}, \bibinfo
  {author} {\bibfnamefont {W.}~\bibnamefont {Eberhardt}}, \bibinfo {author}
  {\bibfnamefont {O.}~\bibnamefont {Rader}}, \ and\ \bibinfo {author}
  {\bibfnamefont {W.}~\bibnamefont {Gudat}},\ }\href@noop {} {\bibfield
  {journal} {\bibinfo  {journal} {Phys. Rev. B}\ }\textbf {\bibinfo {volume}
  {50}},\ \bibinfo {pages} {17496} (\bibinfo {year} {1994})}\BibitemShut
  {NoStop}%
\end{thebibliography}

%
\end{document}